\documentclass[aps,pre,10pt,twocolumn,nolongbibliography,raggedbottom]{revtex4-2}%
\pdfoutput=1
\usepackage{graphicx}%
\usepackage{amsmath, amssymb}%
\usepackage{mathphyscmd}%
\everymath{\displaystyle}%
\usepackage{hyperref}%
\hypersetup{%
	pdfinfo={%
		Author       = {David Gaspard and Jean-Marc Sparenberg},
		Title        = {Multiple scattering model of the quantum random Lorentz gas},
		Subject      = {Scattering theory},
		CreationDate = {D:20210611114545},%
	},
	pdfborderstyle={/S/U/W 1},%
	breaklinks=true
}

\begin{document}%
\title{Multiple scattering model of the quantum random Lorentz gas}%

\author{David Gaspard}
\email[E-mail:~]{dgaspard@ulb.ac.be}
\author{Jean-Marc Sparenberg}
\affiliation{Nuclear Physics and Quantum Physics, CP229, Universit\'e libre de Bruxelles (ULB), École polytechnique, B-1050 Brussels, Belgium}
\date{\today}

\begin{abstract}%
A multiple scattering model of a quantum particle interacting with a random Lorentz gas of fixed point scatterers is established in an Euclidean space of arbitrary dimension.
At the core of the model, the scattering amplitude for the point scatterers is derived in detail, and expressed in terms of the scattering length.
The fundamental properties of the model, such as the cross section and the scattering matrix, are calculated.
In addition, the model is shown to verify the optical theorem and thus probability conservation.
Finally, the differential and total cross sections are numerically computed in two situations whether the Lorentz gas is smaller or larger than the mean free path.
A distinct Airy diffraction peak is obtained for a large enough number of scatterers.
This observation is related to the extinction paradox.
\end{abstract}%
\keywords{Scattering theory; Multiple scattering; Random Lorentz gas; Point scattering; Arbitrary dimension; Optical theorem; Fraunhofer diffraction; Extinction paradox.}
\maketitle

\section{Introduction}\label{sec:introduction}
Multiple scattering theory is a long-standing topic in wave mechanics.
This umbrella term covers several methods of choice to solve wave equations in complex media consisting of a periodic or disordered ensemble of immobile scatterers.
An important one is the Foldy-Lax method~\cite{Foldy1945, Lax1951, *Lax1952, Waterman1961, HuangK2010, Martin2018, Mishchenko2006} which considers point-like scatterers randomly distributed in space.
The scattering of the incident wave is thus described by $s$ waves, because the size of the scatterers is assumed to be small compared to the wavelength.
This kind of model is also referred to as the random Lorentz gas~\cite{Beijeren1982, Kirkpatrick1983a, Kirkpatrick1983b, Esposito1999, Erdos2008, Erdos2007b, Hashimoto2016}.
\par Another important method is the Kohn-Korringa-Rostoker (KKR) method~\cite{Korringa1947, Kohn1954, Korringa1994} which considers non-overlapping spatially extended scatterers.
This method can be regarded as a generalization of the Foldy-Lax method to finite-sized scatterers for which the wave function must be expanded over a partial wave basis on each scattering site.
The KKR method is often used in optics~\cite{Moroz1995}, in solid state physics to compute electronic band structures~\cite{Gonis2000}, but also more fundamentally to study classically chaotic quantum models, such as quantum Sinai billiards~\cite{Berry1981, Wirzba1999}.
\par Foldy's original investigations~\cite{Foldy1945, Lax1951, *Lax1952} led him to develop an integral transport equation for the average intensity of the wave function in the disordered medium.
Nowadays, his equation is well understood in the diagrammatic approach as a consequence of the ladder approximation~\cite{ShengP2006, Akkermans2007, Rossum1999, Kupriyanov2017}.
A corollary of his transport equation is the existence of two different regimes of propagation depending on the size of the system compared to the scattering mean free path: the ballistic regime for small systems and the diffusive regime for large systems~\cite{ShengP2006, Akkermans2007, Beenakker1997, Rossum1999, Kupriyanov2017}.
\par One of the practical advantages of the Foldy-Lax method is that the same formalism can be used regardless of the positions of the scatterers or the number of spatial dimensions. This also includes the one-dimensional case.
In addition, this method avoids the use of large discretization lattices that may not be appropriate at relatively small wavelength.
In the three-dimensional case, similar methods arise in many fields of physics, such as acoustics~\cite{Feuillade1995, Lanoy2015, Martin2018}, seismology~\cite{Zeng1991}, optics~\cite{Berg2008a, *Berg2008b, Pierrat2013, Chabe2014, Bienaime2014} and matter waves~\cite{Kharchenko2001, Champenois2008, Pinsker2016}.
\par In the present paper, we aim at understanding the propagation of a quantum particle of matter, such as an electron, in a disordered medium.
This propagation involves many different processes and, in particular, multiple scattering and diffraction, which are the topic of the present paper.
Our long-term motivation is to model a matter-wave particle evolving in a gaseous particle detector.
\par For this purpose, we first introduce elastic scattering theory in arbitrary dimension.
Although being a relatively straightforward generalization of the well-known 3D scattering theory~\cite{Joachain1979, Newton1982, Taylor2006, Born2019}, this topic is underrepresented in textbooks.
\par Second, we develop an accurate scattering model for the point scatterers, based on the theory of zero-range potentials~\cite{Bolle1984b, Verhaar1985, Albeverio1988, Diejen1991, DeVries1998, Cacciapuoti2007a, *Cacciapuoti2009}.
This theory supersedes the model used in our previous paper~\cite{GaspardD2019b}.
This point scattering model is expressed solely in terms of the scattering length, which is a universal parameter for low-energy scattering~\cite{Newton1982, Joachain1979, Taylor2006, Akkermans2007, Bolle1984b, Verhaar1985, Jeszenszki2018, GaspardD2018a, RamirezSuarez2013, Baye2000a}.
\par Last but not least, we establish the Foldy-Lax method for a random Lorentz gas of point scatterers in a general formulation which does not depend on the number of spatial dimensions.
Particular attention is paid to the applicability of the method to complex values of the wavenumber through analytic continuation, although this question mostly concerns the analytic structure of the Green function.
Based on the tools developed in this paper, we pave the way to a more advanced study of the random Lorentz gas in the complex plane of the wavenumber which continues in our companion paper~\cite{GaspardD2022b}.
\par This paper is organized as follows.
We start by giving a brief introduction to scattering theory in arbitrary dimension in Sec.~\ref{sec:general-scattering}.
This section includes important concepts such as the Green function in Sec.~\ref{sec:free-green-function}, the cross section in Sec.~\ref{sec:general-cross-section}, and the optical theorem in Sec.~\ref{sec:general-optical-theorem}.
\par Then, in Sec.~\ref{sec:point-theory}, we develop a point scattering model intended for the scatterers.
The point scattering amplitude is derived in Secs.~\ref{sec:s-wave-scattering} and~\ref{sec:point-models}, and compared to the delta-like model of Ref.~\cite{Albeverio1988}.
The scattering length is then related to the wave function in Sec.~\ref{sec:scattering-length}, and the cross section is specialized to the point collision in Sec.~\ref{sec:point-cross-section}.
\par The random Lorentz gas model is finally considered in Sec.~\ref{sec:random-lorentz-gas}.
The Foldy-Lax multiple scattering equations are derived in Sec.~\ref{sec:multi-scattering-method}.
They are shown to conserve probability in Sec.~\ref{sec:multi-optical-theorem}.
In addition, a position-space scattering matrix is derived in Sec.~\ref{sec:multi-s-matrix}, and its properties are discussed.
In Sec.~\ref{sec:multi-cross-section}, a numerical study of the differential cross section of the random Lorentz gas is presented in the ballistic and the diffusive regimes. 
\par Since we consider any number of spatial dimensions in this paper, the notations $\mathcal{B}_d$ and $\mathcal{S}_d$ frequently appear.
They respectively refer to the unit $d$-ball and the unit $d$-sphere embedded in the space $\mathbb{R}^d$, such that $\mathcal{S}_d$ is the external border of $\mathcal{B}_d$.
In addition, the corresponding volume and surface area of the unit $d$-ball read
\begin{equation}\label{eq:ball-surf-vol}
V_d = \frac{\pi^{\frac{d}{2}}}{\Gamma(\frac{d}{2}+1)} \qquad\text{and}\qquad
S_d = V_d d = \frac{2\pi^{\frac{d}{2}}}{\Gamma(\frac{d}{2})}  \:,
\end{equation}
where $\Gamma(z)$ denotes the Gamma function~\cite{Olver2010}.
In 3D for instance, this gives $V_3=\tfrac{4\pi}{3}$ and $S_3=4\pi$.

\section{Scattering theory in arbitrary dimension}\label{sec:general-scattering}
In this section, we outline scattering theory in arbitrary dimension, emphasizing fundamental properties such as the scattering amplitude and the cross section.
Although being a relatively straightforward generalization of 3D case, scattering theory in arbitrary dimension is not always presented in textbooks~\cite{Joachain1979, Newton1982, Taylor2006}.
In addition, normalization choices may vary throughout the literature, especially regarding the Green function~\cite{Joachain1979, Newton1982, Taylor2006, Gonis2000, ShengP2006, Akkermans2007, Born2019, Rossum1999}.
Therefore, we believe it is useful to present ours here.

\subsection{Green function in free space}\label{sec:free-green-function}
One considers a spinless particle freely propagating in $d$ spatial dimensions.
The Green function describes, in the stationary picture, the propagation of the particle starting from a source point, and is thus at the core of scattering theory.
We define this function, denoted as $G(k,\vect{r}\mid\vect{r}')$, as the solution of~\cite{Joachain1979, Gonis2000, ShengP2006, Akkermans2007}
\begin{equation}\label{eq:def-free-green}
(\nabla^2 + k^2)G(k,\vect{r}\mid\vect{r}') = \delta^{(d)}(\vect{r}-\vect{r}')  \:,
\end{equation}
where $\textstyle\nabla^2=\sum_{n=1}^d \partial^2_{x_n}$ is the Laplace operator in~$\mathbb{R}^d$, and $\delta^{(d)}(\vect{r}-\vect{r}')$ is the $d$-dimensional Dirac delta function.
\par In Eq.~\eqref{eq:def-free-green}, $k=2\pi/\lambda$ is the wavenumber and $\lambda$ is the wavelength.
It should be noted that this wave equation embeds the dispersion relation between wavenumber and frequency within $k(\omega)$.
In case of a non-relativistic particle, this relation is parabolic: $k(\omega)=(2m\omega/\hbar)^{\frac{1}{2}}$.
If, on the contrary, the particle is relativistic, the relation is hyperbolic: $k(\omega)=((\hbar\omega)^2-(mc^2)^2)^{\frac{1}{2}}/(\hbar c)$.
Therefore, the present considerations may also be extended to electromagnetic waves~\cite{ShengP2006, Akkermans2007, Born2019}.
In this paper, we will consider $k$ as the main variable of the system, instead of the energy $E=\hbar\omega$, in order to avoid loosing generality due to the choice of the dispersion relation.
The same approach is followed in Ref.~\cite{ShengP2006}.
\par
\begin{figure}[ht]%
\includegraphics{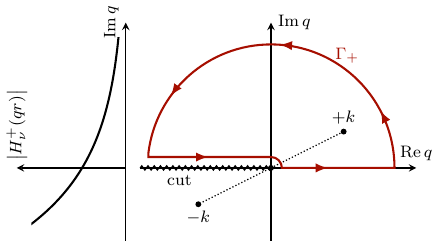}%
\caption{Integration contour used in Eq.~\eqref{eq:free-green-integral-3}.
The left-hand side depicts the exponential behavior of the integrand along the imaginary axis of $q$.}%
\label{fig:green-integral}%
\end{figure}

 The resolvent operator associated with Eq.~\eqref{eq:def-free-green} can be written in Fourier space as~\cite{Joachain1979, Gonis2000, ShengP2006, Akkermans2007}
\begin{equation}\label{eq:def-free-resolvent}
\op{G}(k) = \frac{1}{k^2 - \op{\vect{q}}^2}  \qquad\forall k\in\mathbb{C}\setminus\mathbb{R}  \:,
\end{equation}
where $\op{\vect{q}}=-\I\grad$ denotes the momentum operator. The eigenfunctions of this operator associated with the eigenvalue $\vect{k}$ read
\begin{equation}\label{eq:def-plane-wave}
\braket{\vect{r}}{\vect{k}} = \E^{\I\vect{k}\cdot\vect{r}}  \:.
\end{equation}
In Eq.~\eqref{eq:def-free-resolvent}, the real axis of $k$ is excluded from the domain of the resolvent due to the branch cut singularity, as we will see soon.
The sought Green function is given by the inverse Fourier transform of Eq.~\eqref{eq:def-free-resolvent}~\cite{Joachain1979, Gonis2000, ShengP2006, Akkermans2007}
\begin{equation}\label{eq:free-green-integral-1}
G(k,\vect{r}\mid\vect{r}') = \bra{\vect{r}}\op{G}(k)\ket{\vect{r}'} = \frac{1}{(2\pi)^d} \int_{\mathbb{R}^d} \frac{\E^{\I\vect{q}\cdot(\vect{r}-\vect{r}')}}{k^2 - \vect{q}^2} \D^dq  \:.
\end{equation}
To calculate the integral~\eqref{eq:free-green-integral-1}, we separate the radial and directional parts, and we first integrate over the directional part.
The latter part is given by~\cite{Olver2010}
\begin{equation}\label{eq:plane-wave-spherical-integral}\begin{split}
\oint_{\mathcal{S}_d} \E^{\I z\vect{1}\cdot\vect{\Omega}} \D\Omega & = (2\pi)^{\frac{d}{2}} z^{-\frac{d-2}{2}} J_{\frac{d-2}{2}}(z) \\
 & = S_d\,{}_0F_1(\tfrac{d}{2}, -\tfrac{z^2}{4})  \:,
\end{split}\end{equation}
where $\vect{1}$ is a real unit vector of fixed arbitrary direction, $J_\nu(z)$ is the Bessel function of the first kind, and $\textstyle{}_0F_1(a,z)=\sum_{n=0}^\infty \frac{\Gamma(a)}{\Gamma(a+n)} \frac{z^n}{n!}$ is the generalized hypergeometric function~\cite{Olver2010}.
The second expression of Eq.~\eqref{eq:plane-wave-spherical-integral} better highlights the smooth behavior of the integral at small $z$.
In particular, one can notice that it is an even function of $z$.
Using the notations $r=\norm{\vect{r}-\vect{r}'}$ and $G(k,r)=G(k,\vect{r}\mid\vect{r}')$, the integral~\eqref{eq:free-green-integral-1} becomes
\begin{equation}\label{eq:free-green-integral-2}
G(k,r) = \frac{1}{2\pi} \int_0^\infty \left(\frac{q}{2\pi r}\right)^\frac{d-2}{2} J_{\frac{d-2}{2}}(qr) \frac{q\D q}{k^2 - q^2}  \:.
\end{equation}
The integral path can be closed in the upper complex half-plane of $q$ by separating the two exponential parts of the Bessel function with the property $J_\nu(z)=\left(H^+_\nu(z)+H^-_\nu(z)\right)/2$ in terms of the Hankel functions $H^\pm_\nu(z)$~\cite{Olver2010}.
This leads to the contour integral
\begin{equation}\label{eq:free-green-integral-3}
G(k,r) = \frac{1}{4\pi} \oint_{\Gamma_+} \left(\frac{q}{2\pi r}\right)^\frac{d-2}{2} H^+_{\frac{d-2}{2}}(qr) \frac{q\D q}{k^2 - q^2} \:,
\end{equation}
where $\Gamma_+$ denotes the counter-clockwise half-circle contour running on the real $q$-axis and closing to itself in the upper half-plane of $q$, as shown in Fig.~\ref{fig:green-integral}.
Note that, in contrast to the Bessel function $J_\nu(z)$, the Hankel functions $H^\pm_\nu(z)$ are singular at $z=0$ and possess a branch cut at $\arg z=\pi$ when $\nu$ is an integer. This branch cut is only encountered in even dimensions.
This is why the contour in Fig.~\ref{fig:green-integral} should generally avoid the negative real axis.
Note, in addition, that choosing the symmetric contour with respect to the real $q$ axis and replacing $H^+_\nu(qr)$ by $H^-_\nu(qr)$ does not change the final result of the integral in Eq.~\eqref{eq:free-green-integral-3}.
The integral can then be evaluated with the residue theorem~\cite{Olver2010, Joachain1979, Taylor2006, Akkermans2007} at either $q=+k$ or $q=-k$ depending on which pole is encircled by $\Gamma_+$.
The result is
\begin{equation}\label{eq:resolvent-branch-cut}
G(k,r) = \begin{cases}
G^+(k,r)  & \text{if}~\Im k > 0  \:,\\
G^-(k,r)  & \text{if}~\Im k < 0  \:,
\end{cases}\end{equation}
where the two functions are given by
\begin{equation}\label{eq:free-green-hankel}
G^\pm(k,r) = \pm\frac{1}{4\I} \left(\frac{k}{2\pi r}\right)^\frac{d-2}{2} H^\pm_{\frac{d-2}{2}}(kr)  \:.
\end{equation}
Equation~\eqref{eq:resolvent-branch-cut} confirms the existence of the branch cut of $G(k,r)$ on the real $k$ axis, as already mentioned in Eq.~\eqref{eq:def-free-resolvent}.
This discontinuity means that $G(k,r)$ amounts to either $G^+(k,r)$ or $G^-(k,r)$ depending on the imaginary part, however small, of $k$.
These functions asymptotically behave as~\cite{Olver2010}
\begin{equation}\label{eq:free-green-asym}
G^\pm(k,r) \xrightarrow{r\rightarrow\infty} \pm\frac{1}{2\I k}\left(\frac{\mp\I k}{2\pi r}\right)^{\frac{d-1}{2}} \E^{\pm\I kr} \:,
\end{equation}
and can thus be interpreted as the outgoing Green function with the upper signs, and the incoming Green function with the lower signs.
In the three commonest dimensions, the Green functions read~\cite{Gonis2000, ShengP2006, Akkermans2007}
\begin{equation}
G^\pm(k,r) = \begin{cases}
\E^{\pm\I kr}/(\pm 2\I k)  & \text{for}~d = 1 \:,\\
\pm H^\pm_0(kr)/(4\I)      & \text{for}~d = 2 \:,\\
-\E^{\pm\I kr}/(4\pi r)    & \text{for}~d = 3 \:.
\end{cases}\end{equation}%
\begin{figure}[ht]%
\includegraphics{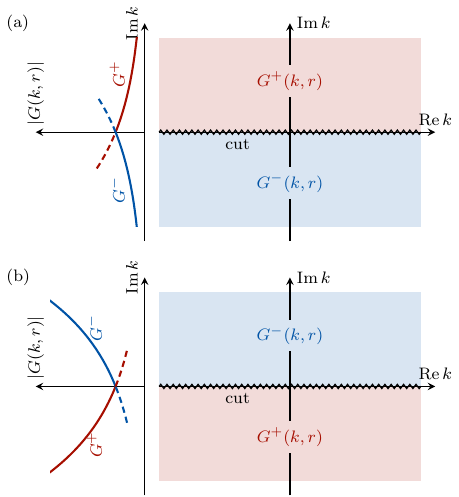}%
\caption{Panel~(a): Structure of the first (physical) Riemann sheet of $G(k,r)$ from Eq.~\eqref{eq:resolvent-branch-cut}.
The left-hand side highlights the exponential behavior of $G^\pm(k,r)$.
Panel~(b): Structure of the second (unphysical) Riemann sheet of $G(k,r)$ obtained by analytic continuation.}%
\label{fig:complex-green}%
\end{figure}%
The structure of the function $G(k,r)$ in Eq.~\eqref{eq:resolvent-branch-cut} is depicted in Fig.~\ref{fig:complex-green}(a) on the first Riemann sheet.
As one can see, $G(k,r)$ displays a branch cut along the real axis of $k$.
It should be noted that $G(k,r)$ is square integrable for all $k\in\mathbb{C}\setminus\mathbb{R}$ because of the exponentially vanishing behavior shown in Eq.~\eqref{eq:free-green-asym}.
This sheet is thus called the physical sheet~\cite{Joachain1979, Newton1982, Taylor2006}.
On the second Riemann sheet, shown in Fig.~\ref{fig:complex-green}(b), the function $G(k,r)$ behaves as an increasing exponential in space, and is thus not square integrable anymore.
This is why the second sheet is referred to as the unphysical sheet~\cite{Joachain1979, Newton1982, Taylor2006}.
This sheet is also considered in scattering theory because it typically contains the resonance poles~\cite{Joachain1979, Newton1982, Taylor2006}.
\par Furthermore, we introduce the auxiliary function $I(k,r)$, which is a regularized version of the Green function at $r=0$.
We define this function for $k\in\mathbb{C}$ as~\cite{Gonis2000, ShengP2006, Akkermans2007}
\begin{equation}\label{eq:def-free-green-imag}
I(k,r) = -\Im[G^+(k,r)] = -\frac{G^+(k,r) - G^-(k,r)}{2\I} \:.
\end{equation}
If one again uses the relation between $H^\pm_\nu(z)$ and $J_\nu(z)$ in Eq.~\eqref{eq:def-free-green-imag}, one finds~\cite{Olver2010}
\begin{equation}\label{eq:free-green-bessel-j}
I(k,r) = \frac{1}{4}\left(\frac{k}{2\pi r}\right)^{\frac{d-2}{2}} J_{\frac{d-2}{2}}(kr)  \:.
\end{equation}
In the three commonest dimensions, this function reads
\begin{equation}
I(k,r) = \begin{cases}
\cos(kr)/(2k)      & \text{for}~d = 1 \:,\\
J_0(kr)/4          & \text{for}~d = 2 \:,\\
\sin(kr)/(4\pi r)  & \text{for}~d = 3 \:.
\end{cases}\end{equation}
Note that, in contrast to $G(k,r)$, the function $I(k,r)$ is an entire function of $r^2$ and regular at $r=0$. The value at this point is
\begin{equation}\label{eq:free-green-imag-zero}
I(k,0) = \frac{\pi}{2}\frac{S_d}{(2\pi)^d}k^{d-2}  \:.
\end{equation}
In addition, an important alternative way of defining the function $I(k,r)$, which better highlights its connection with the density of states (DOS)~\cite{Brennan1999, Gonis2000, Akkermans2007, ShengP2006} in the special case of real values of $k$, is
\begin{equation}\label{eq:def-free-green-imag-ldos}
I(k,\vect{r}\mid\vect{r}') = \pi\bra{\vect{r}}\delta(k^2-\op{\vect{q}}^2)\ket{\vect{r}'} \:.
\end{equation}
The Dirac delta in Eq.~\eqref{eq:def-free-green-imag-ldos} can be interpreted as a projector onto the spherical energy shell
\begin{equation}\label{eq:energy-shell-projector}
\delta(k^2-\op{\vect{q}}^2) = \frac{k^{d-2}}{2(2\pi)^d} \oint_{\mathcal{S}_d} \ket{k\vect{\Omega}}\bra{k\vect{\Omega}}\D\Omega  \:,
\end{equation}
where the plane wave states, $\ket{k\vect{\Omega}}$, are defined by Eq.~\eqref{eq:def-plane-wave}.
Thus, one possible way to the expression~\eqref{eq:def-free-green-imag} of $I(k,r)$ is to directly insert Eq.~\eqref{eq:energy-shell-projector} into Eq.~\eqref{eq:def-free-green-imag-ldos} and then integrate using Eq.~\eqref{eq:plane-wave-spherical-integral}.
Another way is to use the Sokhotski-Plemelj decomposition of the Dirac delta for $k\in\mathbb{R}$~\cite{Vladimirov1971}
\begin{equation}
\delta(k^2-\op{\vect{q}}^2) = \lim_{\varepsilon\downto 0} \left(\frac{(2\pi\I)^{-1}}{k^2-\I\varepsilon - \op{\vect{q}}^2} - \frac{(2\pi\I)^{-1}}{k^2+\I\varepsilon - \op{\vect{q}}^2}\right)  \:,
\end{equation}
which immediately relates Eq.~\eqref{eq:def-free-green-imag-ldos} to the general definition~\eqref{eq:def-free-green-imag} using Eqs.~\eqref{eq:def-free-resolvent} and~\eqref{eq:free-green-integral-1}.
\par It is worth noting that, according to Eq.~\eqref{eq:def-free-green-imag-ldos}, the function $I(k,r)$ is closely related to the DOS per unit of $k^2$ in free space~\cite{ShengP2006, Akkermans2007}
\begin{equation}
\rho_0(k^2) = \Tr[\delta(k^2-\op{\vect{q}}^2)] = \int_{\mathcal{V}} \bra{\vect{r}}\delta(k^2-\op{\vect{q}}^2)\ket{\vect{r}} \D^dr  \:,
\end{equation}
where $\mathcal{V}$ denotes some region of space with volume $V$.
Therefore, one deduces from Eq.~\eqref{eq:def-free-green-imag-ldos} that
\begin{equation}\label{eq:free-green-imag-dos}
\rho_0(k^2) = \frac{V}{\pi} I(k,0)  \:.
\end{equation}
Given that the energy goes as $E\propto k^2$ for a non-relativistic particle, one finds from Eqs.~\eqref{eq:free-green-imag-zero} and~\eqref{eq:free-green-imag-dos} that $\rho_0(E)\propto E^{\frac{d-2}{2}}$, which is the known behavior of the free-space DOS in arbitrary dimension~\cite{Brennan1999, ShengP2006, Akkermans2007}.
\par Finally, we define the real part of the Green function~\eqref{eq:free-green-hankel} as
\begin{equation}\label{eq:def-free-green-real}
P(k,r) = \Re[G^+(k,r)] = \frac{G^+(k,r) + G^-(k,r)}{2}  \:.
\end{equation}
More explicitly, this function reads
\begin{equation}\label{eq:free-green-bessel-y}
P(k,r) = \frac{1}{4}\left(\frac{k}{2\pi r}\right)^{\frac{d-2}{2}} Y_{\frac{d-2}{2}}(kr)  \:,
\end{equation}
where $Y_\nu(z)$ is the Bessel function of the second kind~\cite{Olver2010}.
In this way, the Green function can be fully decomposed into its real and imaginary part as
\begin{equation}\label{eq:free-green-decomposition}
G^\pm(k,r) = P(k,r) \mp\I I(k,r)  \:.
\end{equation}

\subsection{Differential cross section}\label{sec:general-cross-section}
We derive the relation between the scattering amplitude and the cross section in arbitrary dimension.
For this purpose, we consider the Schrödinger equation~\cite{Joachain1979, Newton1982, Gonis2000, ShengP2006, Akkermans2007}
\begin{equation}\label{eq:schrodinger-general}
(\nabla^2 + k^2 - U(\vect{r}))\psi(\vect{r}) = 0  \:,
\end{equation}
where $U(\vect{r})$ denotes a potential function centered at the origin ($\vect{r}=\vect{0}$) with the typical range $R$ such that $U(\vect{r})=0$ for $\norm{\vect{r}}>R$.
We assume that the particle initially enters the collision process in the plane-wave state $\phi(\vect{r})=\braket{\vect{r}}{k\vect{\Omega}_0}=\E^{\I k\vect{\Omega}_0\cdot\vect{r}}$ of incident direction $\vect{\Omega}_0$.
Therefore, in the asymptotic region where the potential $U(\vect{r})$ is zero, the particle wave function $\psi(\vect{r})$ is given by~\cite{Joachain1979, Newton1982, Taylor2006, Gonis2000, ShengP2006, Akkermans2007}
\begin{equation}\label{eq:scat-wave-function-asym}
\psi(\vect{r}) = \phi(\vect{r}) + T(k,\vect{\Omega}) G^+(k,\vect{r}\mid\vect{0})  \:,
\end{equation}
where $\vect{\Omega}=\vect{r}/\norm{\vect{r}}$ denotes the outgoing direction with respect to the incident one.
Expression~\eqref{eq:scat-wave-function-asym} can also be regarded as the practical definition of the scattering amplitude $T(k,\vect{\Omega})$.
Accordingly, the scattering amplitude has the units of length to the power $(d-2)$.
It should be noted that the value of $T(k,\vect{\Omega})$ depends on the normalization of the Green function $G^+(k,r)$.
Due to our definition~\eqref{eq:def-free-green}, $T(k,\vect{\Omega})$ is $-4\pi$ times the standard scattering amplitude in the three-dimensional case~\cite{Joachain1979, Newton1982, Taylor2006, Akkermans2007}.
This choice is motivated by the systematic generalization of scattering theory to other dimensions.
In addition, with this normalization, $T(k,\vect{\Omega})$ coincides with the formal transition operator $\op{T}(k)$~\cite{Joachain1979, Newton1982, Taylor2006, Gonis2000, ShengP2006, Akkermans2007}.
Indeed, they are related by
\begin{equation}\label{eq:def-transition-operator}
T(k,\vect{\Omega}) = \bra{k\vect{\Omega}}\op{T}(k)\ket{k\vect{\Omega}_0}  \:,
\end{equation}
where the plane wave states, $\ket{\vect{k}}$, are defined by Eq.~\eqref{eq:def-plane-wave}.
Of course, this specific normalization has no physical consequence on scattering observables.
\par The differential cross section is usually defined as the ratio between the rate of particle detection in the solid angle $\D\Omega$ and the incoming particle current~\cite{Joachain1979, Newton1982, Akkermans2007, Born2019}, that is to say
\begin{equation}\label{eq:def-diff-cross-section}
\der{\sigma}{\Omega}(k,\vect{\Omega}) = \lim_{r\rightarrow\infty} \frac{\norm{\vect{J}_{\rm out}(\vect{r})}}{\norm{\vect{J}_{\rm in}}}r^{d-1}  \:,
\end{equation}
where the distance $r$ is supposed to be arbitrarily large with respect to the wavelength.
The incoming current of the incident plane wave reads
\begin{equation}\label{eq:incoming-current}
\vect{J}_{\rm in} = \Re[\cc{\phi}(\vect{r})(-\I\grad)\phi(\vect{r})] = k\vect{\Omega}_0  \:.
\end{equation}
Similarly, the outgoing particle current after the collisions is given, according to the wave function~\eqref{eq:scat-wave-function-asym}, by
\begin{equation}\label{eq:outgoing-current-1}
\vect{J}_{\rm out}(\vect{r}) = \abs{T(k,\vect{\Omega})}^2\Re[G^-(k,r)(-\I\grad)G^+(k,r)]  \:.
\end{equation}
The fact that we neglect the angular dependence of the scattering amplitude $T(k,\vect{\Omega})$ when computing the gradient in Eq.~\eqref{eq:outgoing-current-1} comes from the assumption of the far-field regime ($kr\rightarrow+\infty$) that we made in Eq.~\eqref{eq:def-diff-cross-section}.
In this regime, the gradient in Eq.~\eqref{eq:outgoing-current-1} can be approximated by $-\I\grad G^+(k,r)\simeq k\vect{\Omega}G^+(k,r)$, and Eq.~\eqref{eq:outgoing-current-1} becomes
\begin{equation}\label{eq:outgoing-current-2}
\vect{J}_{\rm out}(\vect{r}) \xrightarrow{r\rightarrow\infty} \abs{T(k,\vect{\Omega})}^2 k\vect{\Omega}\abs{G^+(k,r)}^2  \:.
\end{equation}
The function $\abs{G^+(k,r)}^2$ asymptotically behaves as $\bigo(1/r^{d-1})$ according to Eq.~\eqref{eq:free-green-asym}.
In this regard, a useful asymptotic expression can be written, using Eq.~\eqref{eq:free-green-imag-zero}, as
\begin{equation}\label{eq:green-modulus-asym}
\abs{G^+(k,r)}^2 \xrightarrow{r\rightarrow\infty} \frac{I(k,0)}{k S_dr^{d-1}}  \:.
\end{equation}
Then, combining Eqs.~\eqref{eq:incoming-current}, \eqref{eq:outgoing-current-2} and~\eqref{eq:green-modulus-asym} into Eq.~\eqref{eq:def-diff-cross-section}, we obtain the differential cross section~\cite{Joachain1979, Newton1982, Taylor2006, Akkermans2007, ShengP2006}
\begin{equation}\label{eq:diff-cross-section}
\der{\sigma}{\Omega}(k,\vect{\Omega}) = \frac{I(k,0)}{k S_d}\abs{T(k,\vect{\Omega})}^2  \:.
\end{equation}
Note that this expression can be applied to any, however complex, potential $U(\vect{r})$.
This includes either point scatterers or random media, as we will see later.
Finally, the total cross section is given by the integral of Eq.~\eqref{eq:diff-cross-section} over all the outgoing directions~\cite{Joachain1979, Newton1982, Taylor2006, Akkermans2007}
\begin{equation}\label{eq:def-total-cross-section}
\sigma(k) = \oint_{\mathcal{S}_d} \der{\sigma}{\Omega}(k,\vect{\Omega}) \D\Omega  \:.
\end{equation}

\subsection{Optical theorem}\label{sec:general-optical-theorem}
We also derive the probability conservation law, better known in scattering theory as the \emph{optical theorem}~\cite{Joachain1979, Newton1982, Taylor2006, Nussenzveig1992, ShengP2006, Akkermans2007, Born2019, Rossum1999}.
One way is to use the total probability current
\begin{equation}\label{eq:total-current-1}
\vect{J}(\vect{r}) = \Re[\cc{\psi}(\vect{r})(-\I\grad)\psi(\vect{r})]  \:,
\end{equation}
where the wave function $\psi(\vect{r})$ is given by Eq.~\eqref{eq:scat-wave-function-asym}.
Since $\psi(\vect{r})$ contains two terms, the total current~\eqref{eq:total-current-1} splits in three terms: the incident current $\vect{J}_{\rm in}$, the outgoing current $\vect{J}_{\rm out}(\vect{r})$, and the interference current $\vect{J}_{\rm itf}(\vect{r})$. All these contributions are summed up in~\cite{Joachain1979}
\begin{equation}\label{eq:total-current-2}
\vect{J}(\vect{r}) = \vect{J}_{\rm in} + \vect{J}_{\rm out}(\vect{r}) + \vect{J}_{\rm itf}(\vect{r})  \:.
\end{equation}
The probability conservation, $\grad\cdot\vect{J}(\vect{r})=0$, expresses the fact that the spherical flux integral of the total current $\vect{J}(\vect{r})$ around the scattering site amounts to zero. In other words, one has~\cite{Joachain1979}
\begin{equation}\label{eq:total-flux-conservation}
I = \oint_{\mathcal{S}_d} \vect{J}(r\vect{\Omega})\cdot\vect{\Omega} r^{d-1} \D\Omega = I_{\rm in} + I_{\rm out} + I_{\rm itf} = 0  \:,
\end{equation}
where the fluxes $I_{\rm in}$, $I_{\rm out}$ and $I_{\rm itf}$ are defined in the same way as $I$.
In Eq.~\eqref{eq:total-flux-conservation}, we consider the far-field limit ($kr\rightarrow +\infty$) as before, in order to simplify the expressions.
First, the incident term, $I_{\rm in}$, is equal to zero because, due to $\vect{J}_{\rm in}=k\vect{\Omega}_0$, everything that enters the integration sphere ends up leaving it.
Therefore, we may omit this term from now on.
Second, the outgoing term can be related to the total cross section using Eqs.~\eqref{eq:def-diff-cross-section} and~\eqref{eq:def-total-cross-section}
\begin{equation}\label{eq:outgoing-flux}
I_{\rm out} = \oint_{\mathcal{S}_d} \vect{J}_{\rm out}(r\vect{\Omega})\cdot\vect{\Omega} r^{d-1} \D\Omega = k\sigma(k)  \:.
\end{equation}
The only non-trivial term in Eq.~\eqref{eq:total-flux-conservation} is the interference term $I_{\rm itf}$.
This term is the combination of the two cross product terms which appear when inserting Eq.~\eqref{eq:scat-wave-function-asym} into Eq.~\eqref{eq:total-current-1}. It reads
\begin{equation}\label{eq:interference-current}
\vect{J}_{\rm itf}(r\vect{\Omega}) = k(\vect{\Omega} + \vect{\Omega}_0)\Re[\cc{\phi}(\vect{r})T(k,\vect{\Omega})G^+(k,r)]  \:.
\end{equation}
The corresponding flux integral is
\begin{equation}\label{eq:interference-flux-1}\begin{split}
I_{\rm itf} & = \Re\bigg[kG^+(k,r)r^{d-1}  \\
 & \times \oint_{\mathcal{S}_d} (1 + \vect{\Omega}_0\cdot\vect{\Omega})\E^{-\I kr\vect{\Omega}_0\cdot\vect{\Omega}} T(k,\vect{\Omega})\D\Omega\bigg]  \:.
\end{split}\end{equation}
This integral cannot be evaluated exactly for all $T(k,\vect{\Omega})$.
However, we notice that this integral is highly oscillatory in every direction, save in the incident direction $\vect{\Omega}=\vect{\Omega}_0$.
Since the oscillations get faster as $r\rightarrow\infty$, the contribution from all the directions are suppressed except the incident one, so the integrand is proportional to a directional Dirac delta $\delta(\vect{\Omega}-\vect{\Omega}_0)$.
To get the proportionality constant, we consider the same integral but with $T(k,\vect{\Omega})$ replaced by $1$. Using Eq.~\eqref{eq:plane-wave-spherical-integral}, we find
\begin{equation}\label{eq:another-spherical-integral}
\oint_{\mathcal{S}_d} (1 + \vect{\Omega}_0\cdot\vect{\Omega})\E^{-\I kr\vect{\Omega}_0\cdot\vect{\Omega}} \D\Omega
 \xrightarrow{r\rightarrow\infty} -\I S_d \frac{G^-(k,r)}{I(k,0)}  \:.
\end{equation}
With Eq.~\eqref{eq:another-spherical-integral}, expression~\eqref{eq:interference-flux-1} reduces to
\begin{equation}\label{eq:interference-flux-2}
I_{\rm itf} = kS_dr^{d-1} \frac{\abs{G^+(k,r)}^2}{I(k,0)} \Im\left[T(k,\vect{\Omega}_0)\right]  \:.
\end{equation}
This can be further simplified with Eq.~\eqref{eq:green-modulus-asym}. We ultimately obtain
\begin{equation}\label{eq:interference-flux-result}
I_{\rm itf} = \Im[T(k,\vect{\Omega}_0)]  \:.
\end{equation}
Now, inserting Eqs.~\eqref{eq:outgoing-flux} and~\eqref{eq:interference-flux-result} into Eq.~\eqref{eq:total-flux-conservation}, we get the general expression of the optical theorem in arbitrary dimension~\cite{Joachain1979, Newton1982, Taylor2006, Nussenzveig1992, ShengP2006, Akkermans2007, Born2019, Rossum1999}
\begin{equation}\label{eq:general-optical-theorem}
\sigma(k) = -\frac{1}{k}\Im[T(k,\vect{\Omega}_0)]  \:.
\end{equation}
This fundamental result is also very handy to calculate the total cross section without resorting to the integral~\eqref{eq:def-total-cross-section} whose evaluation may be delicate.
We will come back to Eq.~\eqref{eq:general-optical-theorem} several times in this paper.
\par It should be noted that, in the three-dimensional case, the optical theorem is often shown with an additional factor of $-4\pi$ in the right-hand side of Eq.~\eqref{eq:general-optical-theorem}~\cite{Joachain1979, Newton1982, Taylor2006, Nussenzveig1992, Akkermans2007, Born2019, Rossum1999}.
This is again due to our choice of normalization of the free-space Green function $G(k,r)$ in Eq.~\eqref{eq:def-free-green}.
The normalization~\eqref{eq:def-free-green} avoids possibly cumbersome dimension-dependent prefactors in Eq.~\eqref{eq:general-optical-theorem}.

\subsection{One-dimensional case}\label{sec:one-dimension}
Here, we give more insight into the one-dimensional case.
In particular, we show that the cross section is also well defined in 1D, although it is much more rarely used than in higher dimensions~\cite{Markos2008}.
First of all, $\sigma_{\rm 1D}(k)$ is dimensionless, since the cross section is measured in units of length to the power $(d-1)$.
Furthermore, the scattering wave function~\eqref{eq:scat-wave-function-asym} reads
\begin{equation}\label{eq:wave-function-d1}
\psi(x) = \E^{\I kx} + \begin{cases}%
T_+(k) \E^{+\I kx}/(2\I k) & \textrm{if}~x > 0  \:,\\
T_-(k) \E^{-\I kx}/(2\I k) & \textrm{if}~x < 0  \:,
\end{cases}\end{equation}
where $T_+(k)$ and $T_-(k)$ are respectively the forward and backward scattering amplitudes.
In one dimension, it is customary to define the transmission and reflection coefficients, that we denote $A_{\rm T}$ and $A_{\rm R}$, respectively~\cite{Markos2008}.
These coefficients can be related to the scattering amplitudes by
\begin{equation}\label{eq:refl-trans-coef}
A_{\rm T}(k) = 1 + \frac{T_+(k)}{2\I k} \quad\textrm{and}\quad  A_{\rm R}(k) = \frac{T_-(k)}{2\I k}  \:.
\end{equation}
Note the additional unit term in $A_{\rm T}$, coming from the interference between the incident wave $\E^{\I kx}$ and the forwardly scattered wave.
The total cross section $\sigma(k)$ can be obtained from Eqs.~\eqref{eq:diff-cross-section} and~\eqref{eq:def-total-cross-section}
\begin{equation}\label{eq:total-cross-section-d1-1}
\sigma_{\rm 1D}(k) = \frac{\abs{T_+}^2 + \abs{T_-}^2}{(2k)^2} = \abs{A_{\rm T} - 1}^2 + \abs{A_{\rm R}}^2  \:.
\end{equation}
Using the optical theorem~\eqref{eq:general-optical-theorem}, one finds another expression for the total cross section
\begin{equation}\label{eq:total-cross-section-d1-2}
\sigma_{\rm 1D}(k) = -\frac{1}{k}\Im T_+ = 2(1 - \Re A_{\rm T})  \:.
\end{equation}
If one imposes the equality between Eqs.~\eqref{eq:total-cross-section-d1-1} and~\eqref{eq:total-cross-section-d1-2}, one gets a condition on the transmission and reflection coefficients.
After some simplifications, this condition can be written as
\begin{equation}\label{eq:conservation-d1}
\abs{A_{\rm T}}^2 + \abs{A_{\rm R}}^2 = 1  \:.
\end{equation}
This is precisely how probability conservation is supposed to constrain the reflection and transmission coefficients in 1D~\cite{Markos2008}.
This result shows us that the cross section is indeed properly defined in 1D.
This also implies that $A_{\rm T}$ and $A_{\rm R}$ are smaller or equal to unity in absolute value.
A consequence is that, according to Eq.~\eqref{eq:total-cross-section-d1-2}, the one-dimensional cross section possesses an upper bound
\begin{equation}\label{eq:cross-section-bounds-d1}
0 \leq \sigma_{\rm 1D}(k) \leq 4  \:.
\end{equation}
This property is unique to dimension one and has no counterpart in higher dimensions, because the forward scattering amplitude, $T(k,\vect{\Omega}_0)$, is not bounded in general.
The maximum value, $\sigma_{\rm 1D}=4$, corresponds to the case $(A_{\rm T},A_{\rm R})=(-1,0)$, and the minimum value, $\sigma_{\rm 1D}=0$, to $(A_{\rm T},A_{\rm R})=(+1,0)$.
Therefore, in both cases, the scatterer is transparent in the sense that it does not reflect the incident wave.
When the reflection is maximum, the transmission is zero ($A_{\rm T}=0$) and the cross section has the intermediate value $\sigma_{\rm 1D}=2$.
We will come back to Eq.~\eqref{eq:cross-section-bounds-d1} in the framework of point scattering in Sec.~\ref{sec:point-cross-section}.

\subsection{Fraunhofer diffraction}\label{sec:airy-cross-section}
We derive the scattering amplitude and cross section due to a large opaque scatterer of typical radius $R$ in dimension $d\geq 2$.
When the size parameter $kR$ is large, the dominant feature of the scattering amplitude is the forward peak predicted by the Fraunhofer diffraction~\cite{Newton1982, Nussenzveig1992, Hulst1981, Born2019} that we briefly expand here.
We start from the integral form of the Schrödinger equation~\eqref{eq:schrodinger-general} known as the Lippmann-Schwinger equation~\cite{Newton1982, Joachain1979, Taylor2006, Gonis2000}
\begin{equation}\label{eq:airy-lippmann-schwinger}
\psi(\vect{r}) = \E^{\I k\vect{\Omega}_0\cdot\vect{r}} + \int_{\mathbb{R}^d} G^+(k,\vect{r}\mid\vect{x}) U(\vect{x}) \psi(\vect{x}) \D^dx  \:,
\end{equation}
where $G^+(k,\vect{r}\mid\vect{x})$ is the free Green function defined in Eq.~\eqref{eq:def-free-green}, and $U(\vect{x})$ is the potential of the scatterer.
The scattering amplitude $T(k,\vect{\Omega})$ is defined in the Fraunhofer (far-field) regime of the wave function, that is to say $r\gg kR^2/2$.
In this regime, we can use the asymptotic approximation
\begin{equation}\label{eq:fraunhofer-approx}
G^+(k,\vect{r}\mid\vect{x})\xrightarrow{r\rightarrow\infty} G^+(k,\vect{r}\mid\vect{0})\E^{-\I k\vect{\Omega}\cdot\vect{x}}  \:,
\end{equation}
where $\vect{\Omega}=\vect{r}/\norm{\vect{r}}$ is the outgoing direction of the scattering.
The choice of the central position $\vect{r}=\vect{0}$ in the expansion of Eq.~\eqref{eq:fraunhofer-approx} has of course no consequence on the scattering observables.
The scattering angle $\theta$ is defined as the angle with respect to the incident direction, such that $\cos\theta=\vect{\Omega}\cdot\vect{\Omega}_0$.
Then, inserting Eq.~\eqref{eq:fraunhofer-approx} into Eq.~\eqref{eq:airy-lippmann-schwinger}, we can identify the scattering amplitude $T(k,\vect{\Omega})$ defined by Eq.~\eqref{eq:scat-wave-function-asym} as~\cite{Joachain1979, Newton1982}
\begin{equation}\label{eq:scat-ampli-integral}
T(k,\vect{\Omega}) = \int_{\mathbb{R}^d} \E^{-\I k\vect{\Omega}\cdot\vect{x}} U(\vect{x}) \psi(\vect{x}) \D^dx  \:.
\end{equation}
We assume that the potential in Eq.~\eqref{eq:scat-ampli-integral} reads
\begin{equation}\label{eq:airy-ball-potential}
U(\vect{x}) = \begin{cases}
U_0  & \text{for}~\vect{x}\in\mathcal{V} \:,\\ 
0    & \text{otherwise} \:,
\end{cases}\end{equation}
where $\mathcal{V}$ denotes the finite region of $\mathbb{R}^d$ occupied by the scatterer.
Since the scatterer is opaque, the potential~\eqref{eq:airy-ball-potential} tends to infinity ($U_0\rightarrow+\infty$) to obstruct the way of the incident wave.
In Eq.~\eqref{eq:scat-ampli-integral}, we need a convenient approximation for $\psi(\vect{x})~\forall\vect{x}\in\mathcal{V}$ which holds even for such a large potential.
Of course, this excludes the Born approximation which treats the potential $U(\vect{x})$ as a perturbation.
For this purpose, we will resort to a semi-classical, or eikonal, approximation~\cite{Joachain1979, Newton1982, Born2019}, and consider that the incident rays essentially obey a 1D wave equation in the longitudinal direction.
If we decompose the position $\vect{x}$ into the longitudinal and transverse coordinates with respect to the incident direction using
\begin{equation}\label{eq:airy-position-decomposition}
\vect{x} = z\vect{\Omega}_0 + \vect{x}_\perp  \:,
\end{equation}
then the wave function can be approximated as
\begin{equation}\label{eq:eikonal-wave-function-1}
\psi(\vect{x})\simeq\begin{cases}
\E^{\I kz} + A_{\rm R}\E^{-\I k(z-h)}  & \text{if}~z < h(\vect{x}_\perp)    \:,\\
A_{\rm T}\E^{\I\sqrt{k^2-U_0}(z-h)}    & \text{if}~z\geq h(\vect{x}_\perp)  \:,
\end{cases}\end{equation}
where $A_{\rm R}$ and $A_{\rm T}$ are reflection and transmission coefficients, respectively.
In Eq.~\eqref{eq:eikonal-wave-function-1}, the coordinate $z=h(\vect{x}_\perp)$ controls the coordinates of the impact point of the ray on the scatterer surface.
This position still depends on the impact coordinates $\vect{x}_\perp$.
The actual wave function is given by the solution of Eq.~\eqref{eq:eikonal-wave-function-1} for $(A_{\rm R},A_{\rm T})$ based on the continuity conditions at the interface $z=h$.
The result is
\begin{equation}
A_{\rm R} = \frac{k - \sqrt{k^2-U_0}}{k + \sqrt{k^2-U_0}}\E^{\I kh}  \:,\quad
A_{\rm T} = \frac{2k}{k + \sqrt{k^2-U_0}}\E^{\I kh}  \:.
\end{equation}
Therefore, in the limit $U_0\rightarrow+\infty$, the wave function within the scatterer behaves as
\begin{equation}\label{eq:eikonal-wave-function-2}
\psi(\vect{x})\simeq -\frac{2\I k}{\sqrt{U_0}}\E^{\I kh}\E^{-\sqrt{U_0}(z - h)}  \:.
\end{equation}
As a consequence, the product of Eq.~\eqref{eq:eikonal-wave-function-2} with the potential~\eqref{eq:airy-ball-potential} gives rise to a Dirac delta function located on the illuminated face of the scatterer
\begin{equation}\label{eq:dirac-delta-wave-function}
U(\vect{x})\psi(\vect{x}) \xrightarrow{U_0\rightarrow\infty} -2\I k\E^{\I kh} \delta(z - h)  \:.
\end{equation}
If we insert Eq.~\eqref{eq:dirac-delta-wave-function} into Eq.~\eqref{eq:scat-ampli-integral}, we get
\begin{equation}\label{eq:fraunhofer-integral-1}
\frac{T(k,\vect{\Omega})}{-2\I k} = \int_{\mathcal{V}} \E^{-\I k\vect{\Omega}\cdot\vect{x}} \E^{\I kh} \delta(z - h) \D^dx  \:.
\end{equation}
The longitudinal coordinate $z$ can be integrated out with the Dirac delta, leading to
\begin{equation}\label{eq:fraunhofer-integral-2}
\frac{T(k,\vect{\Omega})}{-2\I k} = \int_{\mathcal{S}_\perp} \E^{-\I k\vect{\Omega}\cdot\vect{x}_\perp} \E^{-\I k\vect{\Omega}\cdot\vect{\Omega}_0 h} \E^{\I kh} \D^{d-1}x_\perp  \:,
\end{equation}
where $\mathcal{S}_\perp$ denotes the cross-sectional surface of the scatterer, that is to say the scatterer silhouette.
In Eq.~\eqref{eq:fraunhofer-integral-2}, we notice that $\E^{-\I k\vect{\Omega}\cdot\vect{\Omega}_0 h}\E^{\I kh}=\E^{\I kh(1-\cos\theta)}\simeq 1$ for sufficiently small scattering angles $\theta$.
Therefore, Eq.~\eqref{eq:fraunhofer-integral-2} reduces to
\begin{equation}\label{eq:fraunhofer-integral-3}
\frac{T(k,\vect{\Omega})}{-2\I k} = \int_{\mathcal{S}_\perp} \E^{-\I k\vect{\Omega}\cdot\vect{x}_\perp} \D^{d-1}x_\perp  \:.
\end{equation}
This result highlights the well known property in optics that the Fraunhofer diffraction pattern is given by the Fourier transform of the scatterer silhouette~\cite{Newton1982, Born2019}.
The integral~\eqref{eq:fraunhofer-integral-3} also shows that, in the forward direction ($\vect{\Omega}=\vect{\Omega}_0$), the scattering amplitude is just $T(k,\vect{\Omega}_0)=-2\I k\,S$ where $S$ is the cross-sectional area of the scatterer.
The optical theorem then leads to the total cross section
\begin{equation}\label{eq:fraunhofer-total-cross-section}
\sigma(k) = -\frac{1}{k}\Im[T(k,\vect{\Omega}_0)] = 2S \:,
\end{equation}
which is twice the geometrical cross section of the scatterer, whatever its shape or the number of spatial dimensions.
This may seem paradoxical because, in the semi-classical limit of small wavelengths, one would expect that the total cross section tends to the geometrical cross section without the factor two.
This issue is known as the \emph{extinction paradox}~\cite{Massey1933, Brillouin1949, Hulst1981, Nussenzveig1992, Newton1982, Born2019, Bienaime2014}.
It can be interpreted as a constructive interference effect in the far-field region beyond the shadow cone of the scatterer ($r\gg kR^2/2$).
This subtlety can be traced back from the limit of infinite distance in the definition of the cross section in Eq.~\eqref{eq:def-diff-cross-section}.
To retrieve geometrically admissible cross sections, a possible workaround is to define another cross section with the additional constraint $r\ll kR^2/2$ in Eq.~\eqref{eq:def-diff-cross-section}.
In this way, the interference effect behind the shadow does not contribute to the cross section.
\par In the special case of a spherical scatterer, the silhouette is the disk $\mathcal{S}_\perp=\mathcal{B}_{d-1}(R)$.
It is thus useful to write the integral~\eqref{eq:fraunhofer-integral-3} in spherical coordinates $\vect{x}_\perp=x_\perp\vect{\xi}$, where $\vect{\xi}$ is a unit vector contained in the transverse plane. We can write
\begin{equation}\label{eq:airy-scat-ampli-integral}
\frac{T_{\rm A}(k,\vect{\Omega})}{-2\I k} = \int_0^R \D x_\perp\: x_\perp^{d-2} \oint_{\mathcal{S}_{d-1}} \D\xi \E^{-\I kx_\perp\vect{\Omega}\cdot\vect{\xi}}   \:.
\end{equation}
We define the new unit vector $\vect{\Omega}_\perp$ as the projection of $\vect{\Omega}$ in the transverse plane. 
With this notation, the outgoing direction can be decomposed as
\begin{equation}\label{eq:airy-direction-decomposition}
\vect{\Omega} = \cos\theta\,\vect{\Omega}_0 + \sin\theta\,\vect{\Omega}_\perp  \:,
\end{equation}
and the dot product of Eq.~\eqref{eq:airy-scat-ampli-integral} reads $\vect{\Omega}\cdot\vect{\xi}=\sin\theta\,\vect{\Omega}_\perp\cdot\vect{\xi}$.
Therefore, we identify the inner integral of Eq.~\eqref{eq:airy-scat-ampli-integral} as Eq.~\eqref{eq:plane-wave-spherical-integral}.
After performing the radial integral of Eq.~\eqref{eq:airy-scat-ampli-integral}, we finally obtain the Airy scattering amplitude~\cite{Nussenzveig1992, Born2019}
\begin{equation}\label{eq:airy-scat-ampli-result}
T_{\rm A}(k,\vect{\Omega}) = -2\I k\left(\frac{2\pi R}{k\sin\theta}\right)^\frac{d-1}{2} J_{\frac{d-1}{2}}(kR\sin\theta)  \:.
\end{equation}
Since this formula is valid for small $\theta$, we can write $\sin\theta\simeq\theta$.
We will come back to Eq.~\eqref{eq:airy-scat-ampli-result} in Sec.~\ref{sec:multi-cross-section}.

\section{Point scattering theory}\label{sec:point-theory}
In this section, we establish the scattering amplitude for a single point scatterer.
The topic of point potential has been extensively studied in the literature, especially in Refs.~\cite{Bolle1984b, Verhaar1985, Albeverio1988, Diejen1991, DeVries1998, Cacciapuoti2007a, *Cacciapuoti2009}, but generalizations to an arbitrary number of spatial dimensions are rarely considered.
In this section, we develop a simple $s$-wave scattering model valid in arbitrary dimension inspired by the approach of Ref.~\cite{Verhaar1985}.

\subsection{\texorpdfstring{$s$}{s}-wave scattering}\label{sec:s-wave-scattering}
First of all, it should be noted that $s$-wave models necessarily apply to situations where the wavelength is large compared to the range of the interaction potential, as it is the case, for instance, in the low-energy limit ($k\rightarrow 0$).
These situations also include the zero-range limit of some arbitrarily shaped potential~\cite{Bolle1984b, Verhaar1985}.
Therefore, different calculations will lead to different $s$-wave models with different parameters, the common point being that all the models must behave in the same way in the low-energy limit.
The model derived here is no exception to the rule, and provides the correct low-energy behavior.
In addition, for practical reasons, we parametrize our model in term of the scattering length, which is a commonly used parameter in scattering theory~\cite{Newton1982, Joachain1979, Taylor2006, Akkermans2007, Bolle1984b, Verhaar1985, Jeszenszki2018}.
We will say more about this parameter in Sec.~\ref{sec:scattering-length}.
\par We consider a generic potential $u(\vect{r})$ of finite, and actually small, range~$b$.
The Schrödinger equation reads
\begin{equation}\label{eq:schrodinger-point}
(\nabla^2 + k^2 - u(\vect{r}))\psi(\vect{r}) = 0  \:.
\end{equation}
Since the potential range is assumed to be much smaller than the particle wavelength ($kb\ll 1$), the scattering amplitude in Eq.~\eqref{eq:scat-wave-function-asym} is isotropic and independent of the direction $\vect{\Omega}$~\cite{Joachain1979, Newton1982, Taylor2006}.
To stress this fact, we will use the notation $F(k)$ for the scattering amplitude of the point potential, instead of our general notation $T(k,\vect{\Omega})$ in Eq.~\eqref{eq:scat-wave-function-asym}.
Moreover, it is convenient to define the radial projection of the wave function by integration over the directions
\begin{equation}\label{eq:radial-projection}
\psi(r) = \frac{1}{S_d} \oint_{\mathcal{S}_d} \psi(r\vect{\Omega}) \D\Omega  \:.
\end{equation}
If one applies this radial projection to Eq.~\eqref{eq:scat-wave-function-asym}, then, with the help of Eq.~\eqref{eq:plane-wave-spherical-integral}, one gets
\begin{equation}\label{eq:point-wave-function-1}
\psi(r) = \frac{I(k,r)}{I(k,0)} + F(k) G^+(k,r)  \:.
\end{equation}
The incoming and outgoing parts of this wave function can be separated using Eq.~\eqref{eq:def-free-green-imag}. This leads to
\begin{equation}\label{eq:point-wave-function-2}
\psi(r) = \frac{G^-(k,r) - \left[1 - 2\I I(k,0)F(k)\right] G^+(k,r)}{2\I I(k,0)}  \:.
\end{equation}
In Eq.~\eqref{eq:point-wave-function-2}, one identifies the scattering matrix element between square brackets
\begin{equation}\label{eq:point-s-matrix}
S(k) = \E^{2\I\delta(k)} = 1 - 2\I I(k,0)F(k)  \:.
\end{equation}
The notation $\delta(k)$ stands for the $s$-wave scattering phase shift.
Equation~\eqref{eq:point-s-matrix} shows us that the scattering amplitude $F(k)$ is related to the phase shift by
\begin{equation}\label{eq:point-ampli-from-phase-shift}
F(k)^{-1} = I(k,0)\left(\I - \cot\delta(k)\right)  \:.
\end{equation}
The conservation of probability requires the phase shift $\delta(k)$ to be a real function of $k$, but it does not fix it~\cite{Newton1982, Joachain1979}.
This is the purpose of the next subsection.

\subsection{Determination of the \texorpdfstring{$s$}{s}-wave phase shift}\label{sec:point-models}
In order to determine the phase shift $\delta(k)$ in Eq.~\eqref{eq:point-ampli-from-phase-shift}, we need to solve the Schrödinger equation~\eqref{eq:schrodinger-point} in the inner region of the scatterer potential ($r<b$).
If we assume that the potential $u(\vect{r})$ is spherically symmetric, then Eq.~\eqref{eq:schrodinger-point} can be written in radial coordinates as
\begin{equation}\label{eq:radial-schrodinger}
\left(\pder[2]{}{r} + \frac{d-1}{r}\pder{}{r} + k^2 - u(r)\right)\psi(r) = 0  \:.
\end{equation}
Denoting $\psi(r)$ the solution to Eq.~\eqref{eq:radial-schrodinger} in the inner region, the continuity conditions at the boundary $r=b$ between $\psi(r)$ and the asymptotic function~\eqref{eq:point-wave-function-1} impose that~\cite{Joachain1979, Newton1982}
\begin{equation}\label{eq:point-boundary-condition}
\left.\frac{\partial_r\psi(r)}{\psi(r)}\right|_{r=b} = \left.\frac{\frac{\partial_rI(k,r)}{I(k,0)} + F(k)\partial_rG^+(k,r)}{\frac{I(k,r)}{I(k,0)} + F(k)G^+(k,r)}\right|_{r=b}  \:.
\end{equation}
Note that the logarithmic derivative of $\psi(r)$ at $r=b$ in Eq.~\eqref{eq:point-boundary-condition} is also related to the inverse of the \emph{$R$ matrix} in scattering theory~\cite{Wigner1947, Descouvemont2010}.
Solving Eq.~\eqref{eq:point-boundary-condition} for $F(k)$ leads to
\begin{equation}\label{eq:point-ampli-from-wronskian}
F(k)^{-1} = -I(k,0)\frac{\Wr[G^+(k,r), \psi(r)]_{r=b}}{\Wr[I(k,r), \psi(r)]_{r=b}}  \:,
\end{equation}
where $\Wr[f(r),g(r)]=f(r)\partial_rg(r)-g(r)\partial_rf(r)$ denotes the Wronskian with respect to $r$~\cite{Olver2010}.
Using the decomposition~\eqref{eq:free-green-decomposition} of $G^+(k,r)$ in Eq.~\eqref{eq:point-ampli-from-wronskian}, one may identify the cotangent of the phase shift from Eq.~\eqref{eq:point-ampli-from-phase-shift}. This leads to
\begin{equation}\label{eq:cot-delta-1}
\cot\delta(k) = \frac{\Wr[P(k,r), \psi(r)]_{r=b}}{\Wr[I(k,r), \psi(r)]_{r=b}}  \:.
\end{equation}
The function~\eqref{eq:cot-delta-1} is more suitable to an analysis than Eq.~\eqref{eq:point-ampli-from-wronskian}, because the phase shift is a real function in contrast to $F(k)^{-1}$.
Of course, being related by Eq.~\eqref{eq:point-ampli-from-phase-shift}, both equations are equivalent.
Let us rewrite the Wronskians of Eq.~\eqref{eq:cot-delta-1} in terms of logarithmic derivatives. So, we have
\begin{equation}\label{eq:cot-delta-2}
\cot\delta(k) = \frac{P(k,b)}{I(k,b)} \,\frac{1 - \frac{\psi(b)}{\psi'(b)}\frac{P'(k,b)}{P(k,b)}}{1 - \frac{\psi(b)}{\psi'(b)}\frac{I'(k,b)}{I(k,b)}}  \:,
\end{equation}
where the prime denotes the derivative with respect to~$r$.
In the denominator of Eq.~\eqref{eq:cot-delta-2}, the logarithmic derivative of $I(k,b)$ tends to zero at low energy
\begin{equation}
\frac{I'(k,r)}{I(k,r)} = -\frac{k^2r}{d} + \bigo(k^4r^3)  \qquad\text{for}~k\xrightarrow{>} 0 \:.
\end{equation}
Thus, Eq.~\eqref{eq:cot-delta-2} simplifies to
\begin{equation}\label{eq:cot-delta-3}
\cot\delta(k) = \frac{P(k,b)}{I(k,b)} \underbrace{\left(1 - \frac{\psi(b)}{\psi'(b)}\frac{P'(k,b)}{P(k,b)}\right)}_{A}  \:.
\end{equation}
Two important remarks have to be made about the right hand-side of Eq.~\eqref{eq:cot-delta-3}.
First, it turns out that the prefactor behaves as a power law at small $k$
\begin{equation}\label{eq:cot-delta-power-law}
\frac{P(k,b)}{I(k,b)} = \frac{Y_{\frac{d-2}{2}}(kb)}{J_{\frac{d-2}{2}}(kb)} = \bigo\!\left((kb)^{2-d}\right)  \:.
\end{equation}
Secondly, we notice that the underbraced factor, $A$, behaves as a constant in the same limit.
Therefore, it is legitimate to absorb $A$ into the power law~\eqref{eq:cot-delta-power-law} by replacing the length scale $b$ by $\alpha$.
In this way, Eq.~\eqref{eq:cot-delta-3} becomes
\begin{equation}\label{eq:phase-shift-hard-sphere}
\cot\delta_{\rm hs}(k) = \frac{P(k,\alpha)}{I(k,\alpha)} = \frac{Y_{\frac{d-2}{2}}(\alpha k)}{J_{\frac{d-2}{2}}(\alpha k)}  \:.
\end{equation}
The free real parameter $\alpha$ is known as the \emph{scattering length}~\cite{Newton1982, Joachain1979, Taylor2006, Akkermans2007}.
Our general definition of the scattering length, valid in arbitrary dimension, is consistent with Ref.~\cite{Verhaar1985}.
Inserting Eq.~\eqref{eq:phase-shift-hard-sphere} into Eq.~\eqref{eq:point-ampli-from-phase-shift} leads to the corresponding expression for the scattering amplitude
\begin{equation}\label{eq:point-ampli-hard-sphere}
F_{\rm hs}(k)^{-1} = -I(k,0)\frac{G^+(k,\alpha)}{I(k,\alpha)}  \:.
\end{equation}
Note that this scattering model is only valid for $\alpha k\ll 1$, because it is based on the power law behavior~\eqref{eq:cot-delta-power-law}.
We will explicitly calculate $\alpha$ in Sec.~\ref{sec:scattering-length}.
\par In addition, it should be noted that Eqs.~\eqref{eq:phase-shift-hard-sphere} and thus~\eqref{eq:point-ampli-hard-sphere} can also be obtained in the special case of an infinite potential barrier ($u(r)\rightarrow\infty$).
In this case, the wave function $\psi(b)$ in Eq.~\eqref{eq:cot-delta-3} vanishes, and the scattering length $\alpha$ coincides with $b$.
This shows that $\alpha$ can also be interpreted as the radius of a hard sphere.
This is why we will refer to Eq.~\eqref{eq:phase-shift-hard-sphere} as the \emph{hard-sphere $s$-wave model}, although it actually holds for any potential $u(r)$ under the low-energy assumption.
\par As we explained before, any point scattering model exhibiting the low-energy behavior of Eq.~\eqref{eq:phase-shift-hard-sphere} is a valid model.
Therefore, another obvious scattering model may be derived from the first order expansion of Eq.~\eqref{eq:phase-shift-hard-sphere} at $\alpha k=0$.
This operation gives the \emph{delta-like model} of Ref.~\cite{Albeverio1988}
\begin{equation}\label{eq:phase-shift-delta-like}
\cot\delta_{\rm dl}(k) = \begin{cases}
-\frac{\Gamma(\frac{d-2}{2})\Gamma(\frac{d}{2})}{\pi}\left(\frac{\alpha k}{2}\right)^{2-d}  & \textrm{for}~d\neq 2  \:,\\
\frac{2}{\pi}\left(\ln\!\left(\frac{\alpha k}{2}\right) + \gamma\right)  & \textrm{for}~d = 2  \:,
\end{cases}\end{equation}
where $\gamma=0.57721\ldots$ is the Euler-Mascheroni constant~\cite{Olver2010}.
The name of this model comes from the fact that, if one solves Eq.~\eqref{eq:radial-schrodinger} with the properly renormalized Dirac delta potential of Ref.~\cite{Albeverio1988}, one would obtain the phase shift of Eq.~\eqref{eq:phase-shift-delta-like}.

\subsection{Determination of the scattering length}\label{sec:scattering-length}
Here, we relate the scattering length $\alpha$ to the potential $u(r)$ in Eq.~\eqref{eq:radial-schrodinger} through the wave function $\psi(r)$ which is assumed to be known in the inner region ($r<b$).
In principle, Eq.~\eqref{eq:cot-delta-3} should be compatible with Eq.~\eqref{eq:phase-shift-hard-sphere} in the low-energy limit.
Therefore, imposing the equality between Eqs.~\eqref{eq:cot-delta-3} and~\eqref{eq:phase-shift-hard-sphere}, and writing everything in terms of Bessel functions, we get
\begin{equation}\label{eq:scat-len-eq-1}
\frac{Y_{\frac{d-2}{2}}(\alpha k)}{J_{\frac{d-2}{2}}(\alpha k)} = \frac{Y_{\frac{d-2}{2}}(kb)}{J_{\frac{d-2}{2}}(kb)} \left(1 + \frac{\psi(b)}{\psi'(b)} \frac{k\,Y_{\frac{d}{2}}(kb)}{Y_{\frac{d-2}{2}}(kb)}\right)  \:.
\end{equation}
In order to solve Eq.~\eqref{eq:scat-len-eq-1} for $\alpha$, we have to expand the ratios of Bessel functions in power series for small arguments ($k\rightarrow 0$).
The first ratio of Bessel functions appearing in the left- and right-hand sides of Eq.~\eqref{eq:scat-len-eq-1} behaves as
\begin{equation}\label{eq:bessel-ratio-behavior-1}
\frac{Y_{\frac{d-2}{2}}(z)}{J_{\frac{d-2}{2}}(z)} \xrightarrow{z\rightarrow 0} \begin{cases}
-\frac{\Gamma(\frac{d-2}{2})\Gamma(\frac{d}{2})}{\pi}\left(\frac{z}{2}\right)^{2-d}  & \text{for}~d\neq 2 \:,\\
\frac{2}{\pi}\left(\ln\!\left(\frac{z}{2}\right) + \gamma\right)  & \text{for}~d = 2 \:.
\end{cases}\end{equation}
In addition, the second ratio of Bessel functions in the right-hand side of Eq.~\eqref{eq:scat-len-eq-1} behaves as
\begin{equation}\label{eq:bessel-ratio-behavior-2}
\frac{Y_{\frac{d}{2}}(z)}{Y_{\frac{d-2}{2}}(z)} \xrightarrow{z\rightarrow 0} \begin{cases}
\frac{d-2}{z}  & \text{for}~d\neq 2 \:,\\
-\frac{1}{z\left(\ln(\frac{z}{2}) + \gamma\right)}  & \text{for}~d = 2 \:.
\end{cases}\end{equation}
We notice that, in dimension two, the function $Y_0(z)$ behaves logarithmically for $z\rightarrow 0$, in contrast to the other dimensions.
For this reason, the special case $d=2$ should be treated separately.
Inserting Eqs.~\eqref{eq:bessel-ratio-behavior-1} and~\eqref{eq:bessel-ratio-behavior-2} into Eq.~\eqref{eq:scat-len-eq-1}, and solving for $\alpha$ gives us the general expression of the scattering length in arbitrary dimension
\begin{equation}\label{eq:scat-len-general}
\alpha = \begin{cases}
b\left(1 + (d-2)\frac{\psi(b)}{b\,\psi'(b)}\right)^{-\frac{1}{d-2}}  & \text{for}~d\neq 2  \:,\\
b\,\exp\!\left(-\frac{\psi(b)}{b\,\psi'(b)}\right)  & \text{for}~d=2  \:,
\end{cases}\end{equation}
where the wave function $\psi(r)$ is computed in the limit $k\rightarrow 0$~\cite{Jeszenszki2018, GaspardD2018a, RamirezSuarez2013, Baye2000a}.
Note that the second row of Eq.~\eqref{eq:scat-len-general} is still consistent with the first row in the limit $d\rightarrow 2$. 
This is not trivial because the corresponding rows in Eqs.~\eqref{eq:bessel-ratio-behavior-1} and~\eqref{eq:bessel-ratio-behavior-2} do not apparently match in the limit $d\rightarrow 2$.
\par The formulas in Eq.~\eqref{eq:scat-len-general} can also be retrieved by solving $\psi(\alpha)=0$ for $\alpha$ using one step of a variant of the Newton root-finding method~\cite{Olver2010} based on the ansatz $\psi(r)=c_0+c_1/r^{d-2}$, if $c_0$ and $c_1$ are constants.
This ansatz comes from the general solution of Eq.~\eqref{eq:radial-schrodinger} for $k=0$ in the external region where $u(r)=0$.
In particular, when $d=1$, Eq.~\eqref{eq:scat-len-general} reduces to the standard Newton method.
The appearance of the Newton method highlights the geometrical interpretation of the scattering length as the root of the asymptotic wave function for $k=0$~\cite{Joachain1979}.
\par Finally, let us illustrate the use of Eq.~\eqref{eq:scat-len-general} in the special case of the square well potential
\begin{equation}\label{eq:potential-sqw}
u(r) = \begin{cases}
-w^2  & \text{for}~r < b \:,\\ 
 0    & \text{otherwise} \:.
\end{cases}\end{equation}
In this case, the radial wave function in the inner region reads
\begin{equation}\label{eq:wave-function-sqw}
\psi(r) = I(\sqrt{k^2+w^2},r)  \:,
\end{equation}
where $I(k,r)$ is defined in Eq.~\eqref{eq:free-green-bessel-j}.
Note that this is the only solution to Eq.~\eqref{eq:radial-schrodinger} which is regular at $r=0$.
If, in addition, the well is deep enough ($w\gg k$), then the wave function can be approximated by $\psi(r)=I(w,r)$.
Therefore, knowing the logarithmic derivative
\begin{equation}\label{eq:log-derivative-sqw}
\frac{\psi'(r)}{\psi(r)} = -\frac{w\,J_{\frac{d}{2}}(wr)}{J_{\frac{d-2}{2}}(wr)}  \:,
\end{equation}
then, for this special case, Eq.~\eqref{eq:scat-len-general} becomes
\begin{equation}\label{eq:scat-len-sqw}
\alpha = \begin{cases}
b\left(1 - (d-2)\frac{J_{\frac{d-2}{2}}(wb)}{wb\,J_{\frac{d}{2}}(wb)}\right)^{-\frac{1}{d-2}}  & \text{for}~d\neq 2  \:,\\
b\,\exp\!\left(\frac{J_0(wb)}{wb\,J_1(wb)}\right)  & \text{for}~d=2 \:,
\end{cases}\end{equation}
as given by Eqs.~(37) and~(38) of Ref.~\cite{Verhaar1985}.
These results can also be generalized to a potential barrier, assuming that $w$ in Eq.~\eqref{eq:potential-sqw} is purely imaginary.
This has the effect of replacing the Bessel function $J_\nu(z)$ in Eqs.~\eqref{eq:log-derivative-sqw} and~\eqref{eq:scat-len-sqw} by the modified Bessel function $I_\nu(z)$~\cite{Olver2010}.

\subsection{Cross section of the point scatterer}\label{sec:point-cross-section}
We determine the expression of the total cross section for a single point scatterer, and we show how this cross section must be constrained to satisfy probability conservation.
Since the potential of the scatterer has zero range, the differential cross section~\eqref{eq:diff-cross-section} does not depend on the outgoing direction.
Thus, the integral over the outgoing directions, which gives the total cross section, is immediate
\begin{equation}\label{eq:point-total-cross-section}
\sigma_{\rm pt}(k) = \frac{1}{k}I(k,0) \abs{F(k)}^2  \:.
\end{equation}
Furthermore, we can check that the optical theorem~\eqref{eq:general-optical-theorem} holds for a point scatterer.
Indeed, inserting Eq.~\eqref{eq:point-total-cross-section} into Eq.~\eqref{eq:general-optical-theorem} leads to
\begin{equation}\label{eq:point-optical-theorem}
I(k,0)\abs{F(k)}^2 = -\Im[F(k)]  \:.
\end{equation}
Using the general property $\Im(1/z)=-\Im(z)/\abs{z}^2$, Eq.~\eqref{eq:point-optical-theorem} is equivalent to
\begin{equation}\label{eq:unitary-condition}
\Im[F(k)^{-1}] = I(k,0)  \:.
\end{equation}
Now, we notice that the condition~\eqref{eq:unitary-condition} is always satisfied by Eq.~\eqref{eq:point-ampli-from-phase-shift}, as long as the scattering phase shift $\delta(k)$ is a real number.
This criterion is indeed satisfied by the scattering models~\eqref{eq:phase-shift-hard-sphere} and~\eqref{eq:phase-shift-delta-like}.
\begin{figure}[ht]%
\includegraphics{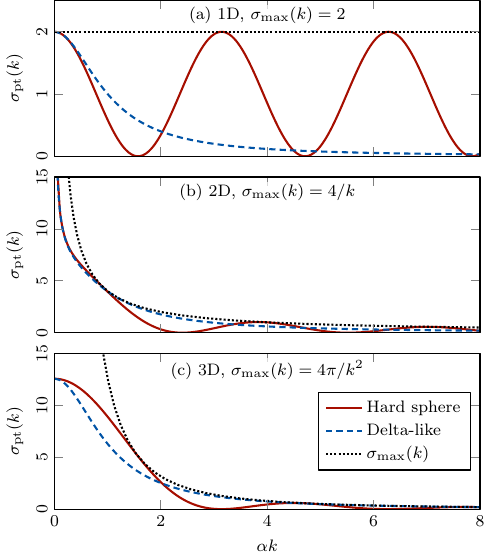}%
\caption{Total cross section~\eqref{eq:point-total-cross-section} of the point scatterer for both the hard-sphere $s$-wave model of Eq.~\eqref{eq:phase-shift-hard-sphere} (solid line), and the delta-like model of Eq.~\eqref{eq:phase-shift-delta-like} (dashed).
The dotted line depicts the upper bound given by Eq.~\eqref{eq:point-cross-section-upper-bound}.
Panels (a)--(c) correspond to the dimensions $d=1,2,3$, respectively.}%
\label{fig:point-cross-section}%
\end{figure}%
\par The total cross sections associated with the models~\eqref{eq:phase-shift-hard-sphere} and~\eqref{eq:phase-shift-delta-like} are shown in Fig.~\ref{fig:point-cross-section}.
The oscillations of the hard-sphere model are due to the function $I(k,\alpha)$ in Eq.~\eqref{eq:point-ampli-hard-sphere}, but are not physically relevant because of the low-energy assumption $\alpha k\ll 1$.
As expected, we notice that the curves indeed match each other in the limit $k\rightarrow 0$.
Moreover, it is remarkable that $d=2$ is the only dimension in which the cross section tends to infinity at zero energy~\cite{Bolle1984b}.
\par Finally, if we insert the $s$-wave scattering amplitude $F(k)$ from Eq.~\eqref{eq:point-ampli-from-phase-shift} into the cross section~\eqref{eq:point-total-cross-section}, we obtain
\begin{equation}\label{eq:point-cross-section-from-phase-shift}
\sigma_{\rm pt}(k) = \frac{\sin^2\delta(k)}{k\,I(k,0)}  \:.
\end{equation}
Since $\sin^2\delta$ is always between 0 and 1, this relation highlights the existence of the following maximum value for the cross section of the point scatterer
\begin{equation}\label{eq:point-cross-section-upper-bound}
\sigma_{\max}(k) = \frac{1}{k\,I(k,0)}  \:.
\end{equation}
The upper bound~\eqref{eq:point-cross-section-upper-bound} decreases with $k$ as $\sigma_{\max}(k)\propto 1/k^{d-1}$.
It is worth noting that this upper bound is universal because it does not depend on the choice of the potential $u(\vect{r})$ or the scattering length~$\alpha$.
However, this only applies to purely isotropic scattering, and can be exceeded if higher order partial waves are added.
As shown in Fig.~\ref{fig:point-cross-section}, the upper bound~\eqref{eq:point-cross-section-upper-bound} gives the correct envelope for both the hard-sphere and the delta-like models.
\par In the one-dimensional case, the upper bound~\eqref{eq:point-cross-section-upper-bound} reduces to $\sigma_{\rm 1D}\leq 2$.
This is more restrictive than our previous bound $\sigma_{\rm 1D}\leq 4$ from Eq.~\eqref{eq:cross-section-bounds-d1}, because of the assumption that the scattering is isotropic in Eq.~\eqref{eq:point-cross-section-upper-bound}.
The bound~\eqref{eq:cross-section-bounds-d1} holds in the more general case of the anisotropic scattering.

\section{Random Lorentz gas model}\label{sec:random-lorentz-gas}
In this section, we consider the case of a particle undergoing elastic collisions without loss of energy in a random Lorentz gas of scatterers in $d$ spatial dimensions~\cite{Esposito1999, Erdos2008, Erdos2007b}.
The wave function $\psi(\vect{r})$ of this particle obeys the stationary Schrödinger equation
\begin{equation}\label{eq:random-gas-schrodinger}
(\nabla^2 + k^2 - U(\vect{r}))\psi(\vect{r}) = 0  \:,
\end{equation}
where the potential $U(\vect{r})$ consists of a sum of $N$ short-range interaction potentials located at the fixed positions $\vect{x}_i~\forall i\in\{1,\ldots,N\}$. It reads
\begin{equation}\label{eq:general-multi-potential}
U(\vect{r}) = \sum_{i=1}^N u(\vect{r}-\vect{x}_i)  \:.
\end{equation}
We suppose that the potential $u(\vect{r})$ has a finite spatial range $b$ much smaller than the wavelength ($kb\ll 1$), so that the point scattering theory of Sec.~\ref{sec:point-theory} can be applied.
In the following subsections, we will use the $s$-wave hard-sphere model of Eq.~\eqref{eq:point-ampli-hard-sphere}, although the main results are not affected by this specific choice.
\par We assume that the positions $\{\vect{x}_1,\vect{x}_2,\ldots,\vect{x}_N\}$ of the scattering sites in Eq.~\eqref{eq:general-multi-potential} are contained in a region $\mathcal{V}$ of finite volume $V$.
These positions are understood as independent and identically distributed random variables uniformly placed in $\mathcal{V}$.
The constant density of scatterers in $\mathcal{V}$ is thus
\begin{equation}\label{eq:def-medium-density}
n = \frac{N}{V} = \frac{1}{\varsigma^d} \:,
\end{equation}
where $\varsigma$ (sigma) is the mean inter-atomic distance. Except otherwise stated, we will treat $\varsigma$ as our unit length.
In addition, we assume that the shape of the Lorentz gas is spherical with the radius
\begin{equation}\label{eq:def-medium-radius}
R = \left(\frac{N}{V_d}\right)^{\frac{1}{d}} \varsigma  \:,
\end{equation}
where $V_d$ is the volume of the unit $d$-ball given by Eq.~\eqref{eq:ball-surf-vol}.
In this way, the density~\eqref{eq:def-medium-density}, and then the unit length $\varsigma$, are kept constant.
A consequence of Eq.~\eqref{eq:def-medium-radius} is that the addition of new scatterers increases the gas size.
The idea is to maintain the gas properties well defined in the limit $N\rightarrow\infty$, in anticipation of future work.
\begin{figure}[ht]%
\includegraphics{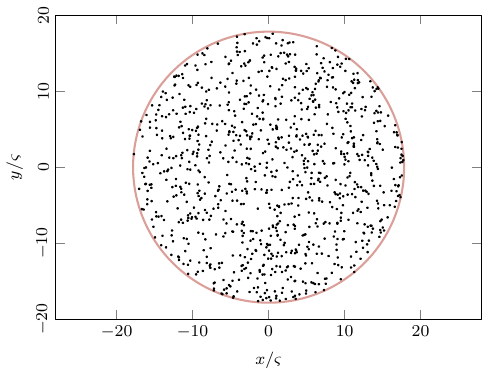}%
\caption{Typical example of a random configuration of independently and uniformly distributed points in a 2D disk for $N=1000$.
The circle highlights the radius of Eq.~\eqref{eq:def-medium-radius}.}%
\label{fig:lorentz-gas-n1000}%
\end{figure}%
\par An example of a random configuration of the scatterers is shown in Fig.~\ref{fig:lorentz-gas-n1000} for $N=1000$ in the two-dimensional case.
According to Eq.~\eqref{eq:def-medium-radius}, the radius is such a gas is $R=(1000/\pi)^{1/2}\varsigma\simeq 17.84\,\varsigma$.

\subsection{Multiple scattering method}\label{sec:multi-scattering-method}
We establish a system of equations to solve the Schrödinger equation~\eqref{eq:random-gas-schrodinger} for the random Lorentz gas.
This method is especially useful for numerical computations, because it avoids the use of heavily discretized methods that may not be appropriate to study the propagation of waves at relatively small wavelength.
\par Given the nature of the potential~\eqref{eq:general-multi-potential}, the particle wave function $\psi(\vect{r})$ should have the form~\cite{Foldy1945, Lax1951, *Lax1952, Mishchenko2006, HuangK2010}
\begin{equation}\label{eq:multi-wave-function}
\psi(\vect{r}) = \phi(\vect{r}) + \sum_{i=1}^N a_i G^+(k,\vect{r}\mid\vect{x}_i)  \:,
\end{equation}
where $\phi(\vect{r})$ is the incident wave, and $a_i~\forall i\in\{1,\ldots,N\}$ is the amplitude scattered by the $i^{\textrm{th}}$ scatterer.
This amplitude is given by the value of the total incident wave function on $\vect{x}_i$, including the waves coming from the other scatterers according to Eq.~\eqref{eq:multi-wave-function}, multiplied by the single-atom scattering amplitude $F(k)$.
Therefore, the self-consistent equation for the amplitudes reads
\begin{equation}\label{eq:lippmann-schwinger}
a_i = F(k)\left(\phi(\vect{x}_i) + \sum_{j (\neq i)}^N a_j G^+(k,\vect{x}_i\mid\vect{x}_j)\right)  \:.
\end{equation}
This equation has the same self-consistent structure as the Lippmann-Schwinger equation that we already encountered in Eq.~\eqref{eq:airy-lippmann-schwinger}, but the role of the wave function is played by the amplitudes $a_i$.
Equation~\eqref{eq:lippmann-schwinger} can be more compactly written in matrix form using the vector notations $\vect{\phi}=\tran{\left(\phi(\vect{x}_1),\phi(\vect{x}_2),\ldots,\phi(\vect{x}_N)\right)}$ and $\vect{a}=\tran{\left(a_1,a_2,\ldots,a_N\right)}$. The result is
\begin{equation}\label{eq:m-matrix-lippmann-schwinger}
\matr{M}(k)\,\vect{a} = \vect{\phi}  \:.
\end{equation}
The $N\times N$ matrix $\matr{M}(k)$ in Eq.~\eqref{eq:m-matrix-lippmann-schwinger} has the elements
\begin{equation}\label{eq:def-m-matrix}
M_{ij}(k) = F(k)^{-1}\delta_{ij} - G^+(k,r_{ij}) (1-\delta_{ij})  \:,
\end{equation}
where $r_{ij}=\norm{\vect{x}_i-\vect{x}_j}$ is the distance between each pair of scatterers.
More explicitly, this matrix has the form
\begin{equation}\label{eq:m-matrix-form}
\matr{M}(k) = \left(\begin{array}{ccc}
F(k)^{-1}       & -G^+(k,r_{1,2}) & \cdots \\
-G^+(k,r_{2,1}) & F(k)^{-1}       & \cdots \\
\vdots          & \vdots          & \ddots
\end{array}\right)  \:.
\end{equation}
Thus, the matrix $\matr{M}(k)$ contains the inverse scattering amplitude along the diagonal, and minus the Green functions between each pair of scatterers off the diagonal.
It is also convenient for the following calculations to introduce the Green matrix $\matr{G}(k)$ with the off-diagonal elements
\begin{equation}\label{eq:def-green-matrix}
G_{ij}(k) = G^+(k,r_{ij}) (1-\delta_{ij})  \:.
\end{equation}
It should be already noted that $\matr{M}(k)$ and $\matr{G}(k)$ are symmetric under matrix transpose due to the fact that $r_{ij}=r_{ji}$.
This can be viewed as a consequence of the time reversal symmetry of the Schrödinger equation~\eqref{eq:random-gas-schrodinger}.
However, $\matr{M}(k)$ and $\matr{G}(k)$ are not Hermitian because the Green function has complex values in general.
Therefore, the eigenvalues of these matrices are also complex.
In addition, $\matr{M}$ is not normal for $N\geq 3$, meaning that it does not commute with its own adjoint
\begin{equation}\label{eq:m-matrix-non-normal}
[\matr{M},\herm{\matr{M}}]\neq 0  \:.
\end{equation}
This somewhat complicates the study of this matrix, as we will see in Sec.~\ref{sec:multi-s-matrix}.
\par We will refer to $\matr{M}(k)$ in Eq.~\eqref{eq:m-matrix-form} as the multiple scattering matrix.
More general matrices of the form $A_{ij}=f(\norm{\vect{x}_i-\vect{x}_j})$ are called Euclidean matrices by some authors~\cite{Skipetrov2011, Goetschy2011a, Goetschy2011b, Goetschy2013-arxiv}.
Furthermore, the method based on Eq.~\eqref{eq:m-matrix-lippmann-schwinger} is often referred to as the Foldy-Lax method~\cite{Foldy1945, Lax1951, *Lax1952, Waterman1961, HuangK2010, Martin2018, Mishchenko2006}.
It is also worth noting that this method can be generalized to describe finite size scatterers by including other partial waves.
If, in addition, appropriate periodic boundary conditions are used, this method is known as the KKR method~\cite{Korringa1947, Kohn1954, Korringa1994}.
It can be successfully applied to the computation of electronic bands in solid state physics~\cite{Gonis2000}, for instance.

\subsection{Cross section and optical theorem}\label{sec:multi-optical-theorem}
We calculate the total cross section of the Lorentz gas, and we establish the relationship with the optical theorem.
As seen before in Sec.~\ref{sec:general-cross-section}, the scattering amplitude $T(k,\vect{\Omega})$ is defined far away from the scattering site by Eq.~\eqref{eq:scat-wave-function-asym}.
Since the Green function asymptotically approaches the Fraunhofer approximation~\eqref{eq:fraunhofer-approx}, that is to say
\begin{equation}\label{eq:shifted-green-asym}
G^+(k,\vect{r}\mid\vect{x}_i) \xrightarrow{r\rightarrow\infty} G^+(k,\vect{r}\mid\vect{0})\E^{-\I k\vect{\Omega}\cdot\vect{x}_i}  \:,
\end{equation}
then the asymptotic behavior of Eq.~\eqref{eq:multi-wave-function} reads
\begin{equation}\label{eq:multi-wave-function-asym}
\psi(\vect{r}) \xrightarrow{r\rightarrow\infty} \phi(\vect{r}) + G^+(k,\vect{r}\mid\vect{0}) \sum_{i=1}^N a_i\E^{-\I k\vect{\Omega}\cdot\vect{x}_i}  \:.
\end{equation}
Comparing with Eq.~\eqref{eq:scat-wave-function-asym}, the sought scattering amplitude can be identified as the sum in the right-hand side of Eq.~\eqref{eq:multi-wave-function-asym}
\begin{equation}\label{eq:multi-scat-ampli-1}
T(k,\vect{\Omega}) = \sum_{i=1}^N a_i \E^{-\I k\vect{\Omega}\cdot\vect{x}_i}  \:.
\end{equation}
It should be noted here that Eq.~\eqref{eq:multi-scat-ampli-1} also gives us the formal transition operator $\op{T}(k)$ of the system.
Indeed, inserting $\vect{a}=\matr{M}^{-1}\vect{\phi}$ in Eq.~\eqref{eq:multi-scat-ampli-1}, we have
\begin{equation}\label{eq:multi-scat-ampli-2}
T(k,\vect{\Omega}) = \sum_{i,j}^N \braket{k\vect{\Omega}}{\vect{x}_i} [\matr{M}^{-1}]_{ij} \braket{\vect{x}_j}{k\vect{\Omega}_0}  \:,
\end{equation}
where the plane wave states, $\ket{\vect{k}}$, are defined by Eq.~\eqref{eq:def-plane-wave}.
Note that Eq.~\eqref{eq:multi-scat-ampli-2} assumes that the incident wave is the momentum eigenstate $\phi(\vect{r})=\braket{\vect{r}}{k\vect{\Omega}_0}$.
Now, identifying $\op{T}(k)$ defined in Eq.~\eqref{eq:def-transition-operator} with Eq.~\eqref{eq:multi-scat-ampli-2}, we get
\begin{equation}\label{eq:multi-transition-operator}
\op{T}(k) = \sum_{i,j}^N [\matr{M}^{-1}]_{ij} \ket{\vect{x}_i}\bra{\vect{x}_j}  \:.
\end{equation}
Beside this, the differential cross section of the whole gas can be derived using Eqs.~\eqref{eq:diff-cross-section} and~\eqref{eq:multi-scat-ampli-1}.
We have
\begin{equation}\label{eq:multi-diff-cross-section}
\der{\sigma}{\Omega}(k,\vect{\Omega}) = \frac{I(k,0)}{kS_d} \sum_{i,j}^N \cc{a}_i a_j \E^{\I k\vect{\Omega}\cdot(\vect{x}_i-\vect{x}_j)}  \:.
\end{equation}
Then, the total cross section can be obtained by integrating Eq.~\eqref{eq:multi-diff-cross-section} over the directions with Eq.~\eqref{eq:plane-wave-spherical-integral} 
\begin{equation}\label{eq:multi-total-cross-section-1}
\sigma(k) = \frac{1}{k}\sum_{i,j}^N \cc{a}_i I(k,r_{ij}) a_j = \frac{1}{k}\herm{\vect{a}}\,\matr{I}(k)\,\vect{a}  \:.
\end{equation}
The new matrix $\matr{I}(k)$ in Eq.~\eqref{eq:multi-total-cross-section-1} is defined as
\begin{equation}\label{eq:def-i-matrix}
I_{ij}(k) = I(k,r_{ij})  \:,
\end{equation}
where $I(k,r)$ is given by Eq.~\eqref{eq:free-green-bessel-j}.
In contrast to $\matr{G}(k)$, the matrix $\matr{I}(k)$ has the nonzero value $I(k,0)$ on the diagonal.
More importantly, one notices that $\matr{I}(k)$ defines a positive-definite quadratic form in Eq.~\eqref{eq:multi-total-cross-section-1} for $k\in\mathbb{R}$.
This is due to the fact that the total cross section is given by the integral of the positive quantity~\eqref{eq:multi-diff-cross-section} and is thus necessarily positive.
A consequence is that the eigenvalues of $\matr{I}(k)$ are all positive.
\par Furthermore, according to the optical theorem~\eqref{eq:general-optical-theorem}, the total cross section is also given by
\begin{equation}\label{eq:multi-total-cross-section-2}
\sigma(k) = -\frac{1}{k}\Im[T(k,\vect{\Omega}_0)] = -\frac{1}{k}\Im[\herm{\vect{\phi}}\,\vect{a}]  \:.
\end{equation}
In Eq.~\eqref{eq:multi-total-cross-section-2}, the scattering amplitude in the forward direction is calculated from Eq.~\eqref{eq:multi-scat-ampli-1} using the fact that $\phi_i=\E^{\I k\vect{\Omega}_0\cdot\vect{x}_i}$.
Substituting Eq.~\eqref{eq:m-matrix-lippmann-schwinger} into Eq.~\eqref{eq:multi-total-cross-section-2}, we get
\begin{equation}\label{eq:multi-total-cross-section-2bis}
\sigma(k) = \frac{1}{k} \herm{\vect{a}}\frac{\matr{M}(k) - \herm{\matr{M}}(k)}{2\I}\vect{a}  \:.
\end{equation}
Then, identifying Eq.~\eqref{eq:multi-total-cross-section-2bis} with Eq.~\eqref{eq:multi-total-cross-section-1} leads to
\begin{equation}\label{eq:def-i-matrix-from-m}
\frac{\matr{M}(k) - \herm{\matr{M}}(k)}{2\I} = \matr{I}(k)  \:.
\end{equation}
Since Eq.~\eqref{eq:def-i-matrix-from-m} is verified for all the matrix elements due to Eq.~\eqref{eq:def-free-green-imag} off the diagonal, and Eq.~\eqref{eq:unitary-condition} on the diagonal, this proves that the optical theorem~\eqref{eq:multi-total-cross-section-2} is valid and that the multiple scattering equations satisfy probability conservation.
\par Another consequence of Eq.~\eqref{eq:def-i-matrix-from-m} is that the eigenvalues of $\matr{M}(k)$ have a positive imaginary part for $k\in\mathbb{R}$.
To see that, we consider the eigendecomposition
\begin{equation}\label{eq:m-matrix-right-eigensystem}
\matr{M}(k)\vect{v}_i = \mu_i\vect{v}_i \qquad\forall i \:.
\end{equation}
If we project both sides onto $\herm{\vect{v}}_i$, and take the imaginary part of the whole, we get
\begin{equation}\label{eq:mu-positive-imag}
\Im\mu_i = \herm{\vect{v}}_i\matr{I}(k)\vect{v}_i > 0  \:.
\end{equation}
Since $\matr{I}$ is positive definite due to Eq.~\eqref{eq:multi-total-cross-section-1}, expression~\eqref{eq:mu-positive-imag} shows that the imaginary parts of the eigenvalues of $\matr{M}$ are positive~\cite{Mitchell2010}.
However, the property~\eqref{eq:mu-positive-imag} does not mean that the imaginary parts $\Im\mu_i$ are related in any way to the eigenvalues of $\matr{I}$ because $[\matr{M},\matr{I}]\neq 0$.
In other words, there is no common eigenbasis for both $\matr{M}$ and $\matr{I}$.

\subsection{Position-space scattering matrix}\label{sec:multi-s-matrix}
Beside the cross section, another fundamental quantity is the scattering operator, $\op{S}(k)$, defined in formal scattering theory as~\cite{Joachain1979, Newton1982, Taylor2006, Akkermans2007}
\begin{equation}\label{eq:def-s-operator}
\op{S}(k) = \op{1} - 2\pi\I\delta(k^2-\op{\vect{q}}^2)\op{T}(k)  \:,
\end{equation}
where the transition operator $\op{T}(k)$ is given by Eq.~\eqref{eq:multi-transition-operator}.
\par Instead of writing $\op{S}(k)$ in the eigenbasis of the free Hamiltonian, i.e., the momentum basis $\ket{\vect{k}}$, as is customary~\cite{Joachain1979, Newton1982, Taylor2006}, the nature of the multiple scattering model suggests to project $\op{S}(k)$ onto the position states of the scatterers, i.e., $\ket{\vect{x}_i}~\forall i\in\{1,\ldots,N\}$.
This leads to a rather unconventional but simpler expression for the scattering operator.
However, care should be taken to the fact that the position states have no finite norm: $\braket{\vect{r}}{\vect{r}'}=\delta(\vect{r}-\vect{r}')$.
The remedy is to define an arbitrarily small volume $B=V_db^d$ for the scatterers, $b$ being their radius.
The orthonormality relation then becomes $\braket{\vect{x}_i}{\vect{x}_j}B=\delta_{ij}$, and we can write
\begin{equation}\label{eq:multi-s-matrix-1}\begin{split}
 & \bra{\vect{x}_i}\op{S}(k)\ket{\vect{x}_j}B = \braket{\vect{x}_i}{\vect{x}_j}B  \\
 & - 2\pi\I\sum_{i'j'}\bra{\vect{x}_i}\delta(k^2-\op{\vect{q}}^2)\ket{\vect{x}_{i'}} [\matr{M}^{-1}]_{i'j'} \braket{\vect{x}_{j'}}{\vect{x}_j}B  \:.
\end{split}\end{equation}
Defining the scattering matrix as $S_{ij}=\bra{\vect{x}_i}\op{S}\ket{\vect{x}_j}B$ and using Eqs.~\eqref{eq:def-free-green-imag-ldos} and~\eqref{eq:def-i-matrix}, we get
\begin{equation}\label{eq:multi-s-matrix-2}
S_{ij}(k) = \delta_{ij} - 2\I\sum_{i'} I_{ii'}(k) [\matr{M}^{-1}]_{i'j}  \:.
\end{equation}
Equation~\eqref{eq:multi-s-matrix-2} can be simplified further with Eq.~\eqref{eq:def-i-matrix-from-m}.
We ultimately obtain the result
\begin{equation}\label{eq:multi-s-matrix}
\matr{S}(k) = \herm{\matr{M}}(k) \matr{M}(k)^{-1}  \:.
\end{equation}
\par It should be noted that, strictly speaking, the scattering matrix~\eqref{eq:multi-s-matrix} is not unitary because neither $\herm{\matr{S}}\matr{S}$ nor $\matr{S}\herm{\matr{S}}$ is equal to the identity matrix.
This is due to the non-normality of the matrix $\matr{M}$ of Eq.~\eqref{eq:m-matrix-non-normal} which is transmitted to $\matr{S}$. 
Despite of this, we can show that the eigenvalues of $\matr{S}$ lie on the unit circle.
To this end, we express Eq.~\eqref{eq:multi-s-matrix} in terms of the decompositions $\matr{M}=\matr{R}+\I\matr{I}$ and $\herm{\matr{M}}=\matr{R}-\I\matr{I}$. We have
\begin{equation}\label{eq:s-matrix-eigensys-1}
\matr{S} = (\matr{R} - \I\matr{I}) (\matr{R} + \I\matr{I})^{-1}  \:.
\end{equation}
Since $\matr{I}$ is symmetric and positive definite as shown in Eq.~\eqref{eq:multi-total-cross-section-1}, it admits the Cholesky factorization $\matr{I}=\matr{L}\tran{\matr{L}}$ for some invertible lower triangular matrix $\matr{L}$ \cite{Golub2013}. So, we can write from Eq.~\eqref{eq:s-matrix-eigensys-1}
\begin{equation}\label{eq:s-matrix-eigensys-2}
\matr{L}^{-1}\matr{S}\matr{L} = \frac{\matr{L}^{-1}\matr{R}\tran{(\matr{L}^{-1})} - \I}{\matr{L}^{-1}\matr{R}\tran{(\matr{L}^{-1})} + \I}  \:.
\end{equation}
We notice that $\matr{L}^{-1}\matr{R}\tran{(\matr{L}^{-1})}$ in Eq.~\eqref{eq:s-matrix-eigensys-2} is a real symmetric matrix because $\tran{\matr{R}}=\matr{R}$. Its eigenvalues are thus real.
Therefore, we deduce that $\matr{S}$ is similar to the manifestly unitary matrix in the right-hand side of Eq.~\eqref{eq:s-matrix-eigensys-2}, and that its eigenvalues lie on the unit circle.
This non-trivial spectral property of $\matr{S}$ would have not hold if $\matr{I}$ was not positive definite, because the eigenvalues of the non-symmetric matrix $\matr{R}\matr{I}^{-1}$ in Eq.~\eqref{eq:s-matrix-eigensys-1} could be complex in general.
This shows that the positive definiteness of $\matr{I}$ is indeed a necessary condition for probability conservation.

\subsection{Numerical differential cross section}\label{sec:multi-cross-section}
In this subsection, we numerically compute the differential cross section of the whole random Lorentz gas from Eq.~\eqref{eq:multi-diff-cross-section} by solving the linear system~\eqref{eq:m-matrix-lippmann-schwinger} with LAPACK~\cite{Anderson1999}.
We consider the scattering observables for both particular configurations of the scatterer positions and averages over many configurations.
We assume that the wavelength is small compared to the mean inter-scatterer distance, but still large compared to the size of the scatterers, so as to remain consistent with the assumptions made in Sec.~\ref{sec:point-theory}.
In this case, two regimes can be observed for the cross section depending on the size of the Lorentz gas compared to the scattering mean free path $\ell=(n\sigma_{\rm pt})^{-1}$.
\par In order to reach the diffusive regime with a limited number of scatterers, we restrict the simulations to the 2D case.
Indeed, according to Eq.~\eqref{eq:def-medium-radius}, the radius of the gas is larger in low dimensions for a fixed density.
The scatterers are thus better exploited in 2D.
Nevertheless, certain results can also be verified in higher dimensions. 

\subsubsection{Ballistic regime}\label{sec:ballistic-regime}
In the first regime, we assume that the gas is small compared to the mean free path, i.e., $n\sigma_{\rm pt}R\ll 1$.
This case is also known as the ballistic regime~\cite{Akkermans2007, ShengP2006, Beenakker1997, Rossum1999, Kupriyanov2017}, and is typically encountered for small systems ($N\lesssim 10$).
In this case, the number of collisions is small enough for the Born approximation to be applied to the inverse of the multiple scattering matrix
\begin{equation}\label{eq:m-matrix-born-expansion}
\matr{M}^{-1} = (F^{-1} - \matr{G})^{-1} = F + F^2\matr{G} + F^3\matr{G}^2 + \cdots  \:.
\end{equation}
This means that $\vect{a}\simeq F\vect{\phi}$ at the first order of perturbations in Eq.~\eqref{eq:m-matrix-born-expansion}.
In other words, the amplitudes are approximately proportional to the incident wave function $\phi(\vect{r})$.
Therefore, using the incident wave $\phi(\vect{r})=\E^{\I k\vect{\Omega}_0\cdot\vect{r}}$, the configurational average of the square modulus of the scattering amplitude~\eqref{eq:multi-scat-ampli-1} reads
\begin{equation}\label{eq:avg-t2-born-1}
\avg{\abs{T(k,\vect{\Omega})}^2} \simeq \abs{F(k)}^2 \sum_{i,j}^N \avg{\E^{\I k\Delta\vect{\Omega}\cdot(\vect{x}_i-\vect{x}_j)}}  \:.
\end{equation}
where $\Delta\vect{\Omega}=\vect{\Omega}-\vect{\Omega}_0$.
The sum in Eq.~\eqref{eq:avg-t2-born-1} contains two kinds of term: the diagonal terms ($i=j$) which are just $1$, and the off-diagonal terms ($i\neq j$).
We explicitly separate these terms as
\begin{equation}\label{eq:avg-t2-born-2}
\avg{\abs{T(k,\vect{\Omega})}^2} \simeq \abs{F(k)}^2 \left[N + N(N-1)c(k,\vect{\Omega})\right] \:,
\end{equation}
where $c(k,\vect{\Omega})$ denotes the average of a single off-diagonal term in Eq.~\eqref{eq:avg-t2-born-1}, that we define as
\begin{equation}\begin{split}\label{eq:avg-t2-born-3}
c(k,\vect{\Omega}) & = \avg{\E^{\I k\Delta\vect{\Omega}\cdot(\vect{x}_1-\vect{x}_2)}}  \\
 & = \frac{1}{V^2} \iint_{\mathcal{V}} \E^{\I k\Delta\vect{\Omega}\cdot(\vect{x}_1-\vect{x}_2)} \D^dx_1 \D^dx_2 \:.
\end{split}\end{equation}
Indeed, it should be noted that the scatterer positions are all independently distributed.
Therefore, the off-diagonal terms in Eq.~\eqref{eq:avg-t2-born-1} are identical, and then we may consider a specific pair $(\vect{x}_1,\vect{x}_2)$ to do the calculation in Eq.~\eqref{eq:avg-t2-born-3}.
The factor $N(N-1)$ in Eq.~\eqref{eq:avg-t2-born-2} accounts for the number of identical off-diagonal terms in Eq.~\eqref{eq:avg-t2-born-1}.
\par In addition, the function $c(k,\vect{\Omega})$ can also be calculated for a ball-shaped medium using the integral~\cite{Olver2010}
\begin{equation}\label{eq:plane-wave-ball-integral}
\frac{1}{V} \int_{\mathcal{B}_d(R)} \E^{\I\vect{q}\cdot\vect{r}} \D^dr = \Gamma(\tfrac{d+2}{2}) \left(\frac{2}{qR}\right)^{\frac{d}{2}} J_{\frac{d}{2}}(qR) \:,
\end{equation}
where $V=V_dR^d$ is the volume of the ball-shaped Lorentz gas, and $q=\norm{\vect{q}}$ is the Fourier variable.
In the context of Eq.~\eqref{eq:avg-t2-born-3}, this variable is related to the scattering angle $\theta$ by~\cite{Joachain1979, Newton1982}
\begin{equation}\label{eq:transferred-momentum}
q = k\norm{\Delta\vect{\Omega}} = 2k\sin(\theta/2)  \:.
\end{equation}
The sought function of Eq.~\eqref{eq:avg-t2-born-3} thus reads
\begin{equation}\label{eq:avg-t2-born-4}
c(k,\vect{\Omega}) = \left[\Gamma(\tfrac{d+2}{2}) \left(\frac{2}{qR}\right)^{\frac{d}{2}} J_{\frac{d}{2}}(qR)\right]^2  \:.
\end{equation}
Using Eqs.~\eqref{eq:diff-cross-section} and~\eqref{eq:point-total-cross-section}, the average~\eqref{eq:avg-t2-born-2} leads to the approximate differential cross section
\begin{equation}\label{eq:diff-cross-section-born}
\der{\sigma}{\Omega}(k,\vect{\Omega}) \simeq N\frac{\sigma_{\rm pt}(k)}{S_d} \left[1 + (N-1)c(k,\vect{\Omega})\right]   \:.
\end{equation}%
\begin{figure}[ht]%
\includegraphics{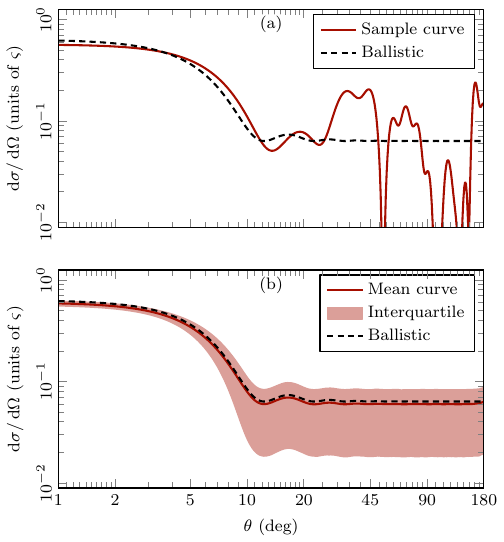}%
\caption{Differential cross section of a 2D disk-shaped Lorentz gas for $N=10$ and $k=10\,\varsigma^{-1}$, using the model of Eq.~\eqref{eq:point-ampli-hard-sphere} with $\alpha=10^{-3}\,\varsigma$.
Panel~(a): Sample curve for a single random configuration of the scatterer positions.
Panel~(b): Average over $2^{18}$ random configurations. The filled region is the interquartile range.
The dashed curve depicts the ballistic approximation~\eqref{eq:diff-cross-section-born} with $c(k,\vect{\Omega})$ from Eq.~\eqref{eq:avg-t2-born-4}.}%
\label{fig:diff-cross-section-ballistic}%
\end{figure}%
\par The approximation~\eqref{eq:diff-cross-section-born} is compared to the actual differential cross section numerically computed from Eq.~\eqref{eq:multi-diff-cross-section} in Fig.~\ref{fig:diff-cross-section-ballistic} for the two-dimensional case.
Very similar curves can be obtained in higher dimensions too.
The cross section of a single scatterer in Fig.~\ref{fig:diff-cross-section-ballistic} is $\sigma_{\rm pt}\simeq 0.0399\,\varsigma$, and the scale parameter is thus $n\sigma_{\rm pt}R\simeq 0.0711<1$.
\par The solid curve in Fig.~\ref{fig:diff-cross-section-ballistic}(a) depicts the differential cross section for a single configuration of the scatterers.
This curve displays a forward peak coming from the term $c(k,\vect{\Omega})$ in Eq.~\eqref{eq:diff-cross-section-born}.
This peak can also be interpreted as a constructive interference between the scatterers due to the variation of the complex phase of the amplitudes $a_i\simeq F(k)\phi(\vect{x}_i)$.
The width of the forward peak can be estimated from the first zero of Eq.~\eqref{eq:avg-t2-born-4}, that is to say
\begin{equation}\label{eq:peak-width-ballistic}
\theta_0 = \frac{j_{\frac{d}{2}}}{kR}  \:,
\end{equation}
where $j_\nu$ denotes the first zero of the Bessel function $J_\nu(z)$.
The first useful values are $j_1=3.83\ldots$ and $j_{\frac{3}{2}}=4.49\ldots$.
According to Eq.~\eqref{eq:peak-width-ballistic}, the width of the peak in Fig.~\ref{fig:diff-cross-section-ballistic} is $\theta_0=12.3^\circ$.
\par Next to the forward peak, the solid curve in Fig.~\ref{fig:diff-cross-section-ballistic}(a) presents fluctuations which depend on the position of the scatterers but also the incident direction.
To determine the characteristic angular scale of these fluctuations, we briefly focus on two scatterers separated by the maximum possible distance $2R$.
Without loss of generality, the scatterers can be placed on the $z$ axis.
The differential cross section of these two isolated scatterers is roughly given by
\begin{equation}\label{eq:diff-cross-section-two-atoms}
\der{\sigma}{\Omega}(k,\theta) \propto \cos^2(kR\cos\theta)  \:,
\end{equation}
where $\theta$ is the polar angle between the observation point and the $z$ axis.
The fastest oscillations of Eq.~\eqref{eq:diff-cross-section-two-atoms} are located in the plane transverse to the $z$ axis, i.e., in the vicinity of $\theta=\tfrac{\pi}{2}$.
The period of the oscillations in this region is
\begin{equation}\label{eq:fluctuations-period}
\Delta\theta = \frac{\pi}{kR}  \:.
\end{equation}
Equation~\eqref{eq:fluctuations-period} gives a plausible estimate of the angular scale of the fluctuations in Fig.~\ref{fig:diff-cross-section-ballistic}(a).
It is also consistent with the period of the angular oscillations of the function $c(k,\vect{\Omega})$ in Eq.~\eqref{eq:avg-t2-born-4}, and always of the same order of magnitude as Eq.~\eqref{eq:peak-width-ballistic}.
\par In Fig.~\ref{fig:diff-cross-section-ballistic}(b), the differential cross section of Eq.~\eqref{eq:multi-diff-cross-section} is numerically averaged over a large number of configurations.
The statistical dispersion of the differential cross section around the mean curve is measured by the interquartile range which is computed for each given value of $\theta$.
This range is defined as the interval between the first and the third quartile of a statistical distribution, and thus contains 50\% of the samples~\cite{Ross2010}.
This range gives more insight into the statistical distribution than the standard deviation, as it also reveals the possible asymmetries.
The large dispersion in Fig.~\ref{fig:diff-cross-section-ballistic}(b) also confirms the sensitivity of the differential cross section to the configuration of the scatterers for $\theta\gtrsim\theta_0$.
\par Finally, we consider the total cross section of the Lorentz gas in the ballistic regime.
The total cross section can be obtained from the integral of Eq.~\eqref{eq:diff-cross-section-born} over the direction.
The result can be written as
\begin{equation}\label{eq:total-cross-section-born}
\avg{\sigma(k)} = N\sigma_{\rm pt}(k) \left[1 + (N-1)C(k)\right] \:,
\end{equation}
where the function $C(k)$ is given for a spherical medium by
\begin{equation}\begin{split}\label{eq:c-integral}
C(k) & = \frac{1}{S_d} \oint_{\mathcal{S}_d} c(k,\vect{\Omega}) \D\Omega \\
 & = {}_2F_3\!\left(\substack{\frac{d-1}{2},\,\frac{d+1}{2}\\ \frac{d+2}{2},\, d-1,\, d+1}; -4(kR)^2\right)  \:.
\end{split}\end{equation}
In Eq.~\eqref{eq:c-integral}, ${}_pF_q(a_1,\ldots,a_p;b_1,\ldots,b_q;z)$ denotes the generalized hypergeometric function~\cite{Olver2010}.
Note that, since $c(k,\vect{\Omega})=1$ for $k=0$, one has the property $C(0)=1$.
Moreover, for large $kR$, $C(k)$ decreases as the power law~\cite{Olver2010}
\begin{equation}\label{eq:c-integral-asym}
C(k) \xrightarrow{kR\rightarrow\infty} \frac{2^d \Gamma(\frac{d}{2})\Gamma(\frac{d+2}{2})^2}{\pi^{3/2}\Gamma(\frac{d+3}{2})} \frac{1}{(kR)^{d-1}} \:.
\end{equation}
Therefore, in the limit $kR\rightarrow\infty$ and for fixed $N$, the total cross section~\eqref{eq:total-cross-section-born} tends to
\begin{equation}\label{eq:total-cross-section-additive}
\avg{\sigma(k)} \simeq N\sigma_{\rm pt}(k)  \:,
\end{equation}
which is $N$ times the point cross section $\sigma_{\rm pt}$ of Eq.~\eqref{eq:point-total-cross-section}.
This means that, in the ballistic regime, all the collisions are mainly independent of each other, and the total cross section is the sum of the individual cross sections of the scatterers.
This property is known as the additive approximation of the ballistic regime~\cite{Hulst1981, Newton1982, Berg2008b, Martin2018}.
We also numerically observed this property in our previous paper~\cite{GaspardD2019b} for $N\leq 10$ in a many-channel version of the present model.
\par Remarkably, if we had used the optical theorem~\eqref{eq:multi-total-cross-section-2} in conjunction with $a_i\simeq F(k)\phi(\vect{x}_i)$ to determine the total cross section, we would have directly found the additive approximation~\eqref{eq:total-cross-section-additive} without going through the expected result of Eq.~\eqref{eq:total-cross-section-born}.
This comes from the fact that the first-order perturbative approximation of Eq.~\eqref{eq:m-matrix-born-expansion} does not conserve probability, thus the optical theorem is not supposed to hold anymore, and Eq.~\eqref{eq:total-cross-section-born} cannot be obtained in this way.
\par
\begin{figure}[ht]%
\includegraphics{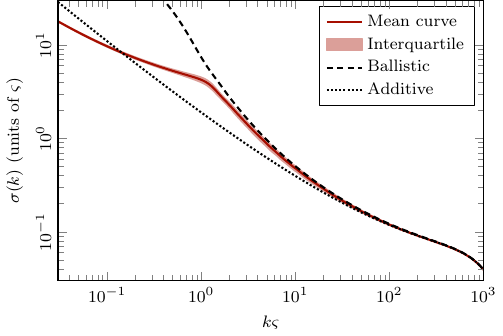}%
\caption{Total cross section of a 2D disk-shaped Lorentz gas for $N=10$, using the model of Eq.~\eqref{eq:point-ampli-hard-sphere} with $\alpha=10^{-3}\,\varsigma$, and averaging over $2^{12}$ random configurations.
The dashed curve is the ballistic approximation~\eqref{eq:total-cross-section-born} with $C(k)$ from Eq.~\eqref{eq:c-integral}, and the dotted curve is the additive approximation~\eqref{eq:total-cross-section-additive}.}%
\label{fig:total-cross-section-ballistic}%
\end{figure}

 The approximations~\eqref{eq:total-cross-section-born} and~\eqref{eq:total-cross-section-additive} are compared to the numerical total cross section of the Lorentz gas in Fig.~\ref{fig:total-cross-section-ballistic}.
As expected, the approximation~\eqref{eq:total-cross-section-born} better matches the exact cross section (in solid) than Eq.~\eqref{eq:total-cross-section-additive}, and seems valid until $k\approx 1\,\varsigma^{-1}$.
Below this point, the cross section suddenly changes. This could be explained by the combination of two facts.
First, the single-scatterer cross section tends to infinity in this region, as shown in Fig.~\ref{fig:point-cross-section}(b), breaking down the validity of the Born approximation.
Second, the wavelength becomes much larger than the interscatterer distance, and one may expect that the multiple scattering be collective in this case.
The vague peak near $k=1\,\varsigma^{-1}$ results from the closeness of the resonance band which is highlighted in our companion paper~\cite{GaspardD2022b}.
\par Finally, note that the small inflection in Fig.~\ref{fig:total-cross-section-ballistic} near $k\simeq 10^3\,\varsigma^{-1}$ is due to the single-scatterer cross section.
Indeed, with the $s$-wave hard-sphere model of Eq.~\eqref{eq:point-ampli-hard-sphere}, the point cross section vanishes at $\alpha k=j_{\frac{d-2}{2}}$, which corresponds to $k\simeq 2405\,\varsigma^{-1}$ in Fig.~\ref{fig:total-cross-section-ballistic}.

\subsubsection{Diffusive regime}\label{sec:diffusive-regime}
Now, we consider the opposite situation when the gas is much larger than the mean free path ($n\sigma_{\rm pt}R\gg 1$).
In this case, the number of collisions becomes large, and the Born approximation does not hold anymore.
This situation is also known as the diffusive regime in reference to the classical Lorentz gas model for which the diffusion equation is expected to hold~\cite{Akkermans2007, ShengP2006, Beenakker1997, Rossum1999, Kupriyanov2017}.
Since the whole gas obstructs the path of the incident particle in the forward direction, it is legitimate to approach the forward scattering amplitude with the model of large opaque sphere of Sec.~\ref{sec:airy-cross-section}.
The forward scattering amplitude of such an obstacle is given by the Airy diffraction pattern~\eqref{eq:airy-scat-ampli-result}
\begin{equation}\label{eq:airy-scat-ampli}
T_{\rm A}(k,\theta) = -2\I k\left(\frac{2\pi R}{k\theta}\right)^{\frac{d-1}{2}} J_{\frac{d-1}{2}}(kR\theta)  \:.
\end{equation}
At small angles ($kR\theta\ll 1$), the main peak of the pattern can be approached by
\begin{equation}\label{eq:airy-scat-ampli-expansion}
T_{\rm A}(k,\theta) = -2\I k V_{d-1}R^{d-1} \left(1 - \frac{(kR\theta)^2}{2(d+1)} + \bigo(\theta^4)\right)  \:.
\end{equation}
This expansion shows that the Airy pattern becomes larger and sharper in $\theta$ as the product $kR$ increases.
According to Eq.~\eqref{eq:diff-cross-section}, the Airy differential cross section corresponding to Eq.~\eqref{eq:airy-scat-ampli} is
\begin{equation}\label{eq:airy-diff-cross-section}
\der{\sigma_{\rm A}}{\Omega}(\theta) = \left[\left(\frac{R}{\theta}\right)^{\frac{d-1}{2}} J_{\frac{d-1}{2}}(kR\theta)\right]^2  \:.
\end{equation}%
\begin{figure}[ht]%
\includegraphics{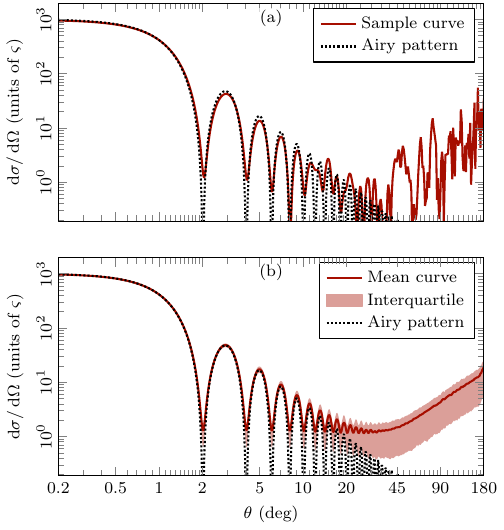}%
\caption{Differential cross section of a 2D disk-shaped Lorentz gas for $N=1000$ and $k=5\,\varsigma^{-1}$, using the point scattering model~\eqref{eq:point-ampli-hard-sphere} with $\alpha=0.1\,\varsigma$.
Panel~(a): Sample curve for a single random configuration of the scatterer positions.
Panel~(b): Average over $2^{12}$ random configurations. The dotted curve depicts the Airy diffraction pattern~\eqref{eq:airy-diff-cross-section}.}%
\label{fig:diff-cross-section-airy}%
\end{figure}%
\par The Airy cross section~\eqref{eq:airy-diff-cross-section} is graphically compared to the actual cross section~\eqref{eq:multi-diff-cross-section} of the Lorentz gas in Fig.~\ref{fig:diff-cross-section-airy}.
Very similar curves are obtained in other dimensions $d>2$.
In Fig.~\ref{fig:diff-cross-section-airy}, the gas contains $N=1000$ scatterers of individual cross section $\sigma_{\rm pt}(k)\simeq 0.65\,\varsigma$.
The scale parameter is thus $n\sigma_{\rm pt}R\simeq 11.6>1$.
\par The sample function shown in Fig.~\ref{fig:diff-cross-section-airy}(a) fluctuates on the angular scale $\Delta\theta=\pi/(kR)$, which is the same formula as Eq.~\eqref{eq:fluctuations-period} for the ballistic regime.
Indeed, as shown in Eq.~\eqref{eq:diff-cross-section-two-atoms}, this is the smallest oscillation period possible for the differential cross section, and it does not depend on the number of scatterers involved in the collisions.
\par The averaged curves are shown in Fig.~\ref{fig:diff-cross-section-airy}(b).
As expected, the Airy pattern does not describe the differential cross section at large scattering angle ($\theta\gtrsim 20^\circ$).
In this region, the cross section increases on average with $\theta$, in contrast to the constant behavior observed for the ballistic case in Fig.~\ref{fig:diff-cross-section-ballistic}.
This could likely be explained through a semi-classical approximation~\cite{Foldy1945, Lax1951, *Lax1952, Mishchenko2006, Akkermans2007, ShengP2006, Akkermans1986, Rossum1999, Chabe2014, Kupriyanov2017}, but this is beyond the scope of this paper.
\par In the backward direction ($\theta=180^\circ$), we notice a faint peak that we interpret as the coherent backscattering peak~\cite{Mishchenko2006, Akkermans2007, ShengP2006, Akkermans1986, Tiggelen1990, Lagendijk1996, Chabe2014, Kupriyanov2017}.
Indeed, the width coincides with the expected angular scale $\Delta\theta_{\rm bs}=1/(k\ell)$, which is about $7.5^\circ$ in Fig.~\ref{fig:diff-cross-section-airy}.
\par Regarding the total cross section, the optical theorem~\eqref{eq:general-optical-theorem} allows us to get a reasonable estimate.
Using Eq.~\eqref{eq:airy-scat-ampli-expansion}, the total scattering cross section is given by
\begin{equation}\label{eq:airy-total-cross-section}
\sigma = -\frac{1}{k}\Im[T_{\rm A}(0)] = 2V_{d-1}R^{d-1}  \:,
\end{equation}
or more specifically, $\sigma=4R$, for a 2D Lorentz gas.
The result~\eqref{eq:airy-total-cross-section} is twice the geometrical cross section of the gas $\sigma_{\rm geom}=V_{d-1}R^{d-1}$.
This is the famous extinction paradox~\cite{Massey1933, Brillouin1949, Hulst1981, Nussenzveig1992, Newton1982, Born2019, Bienaime2014}, already presented in Sec.~\ref{sec:airy-cross-section}.
This phenomenon has never been highlighted before for a random Lorentz gas of point scatterers in arbitrary dimension.
However, its involvement in the context of multiple scattering is somewhat fortuitous because it does not rely on the disordered structure of the system but only the fact that the wave is strongly scattered in all directions save the incident one.
\begin{figure}[ht]%
\includegraphics{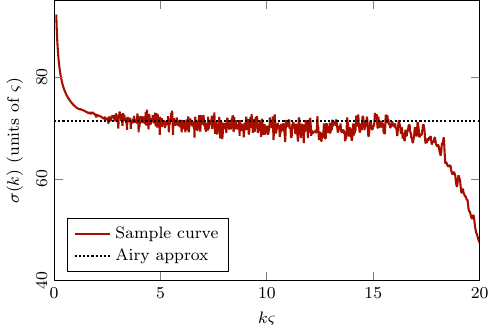}%
\caption{Total cross section of a 2D disk-shaped Lorentz gas for $N=1000$, using the model of Eq.~\eqref{eq:point-ampli-hard-sphere} with $\alpha=0.1\,\varsigma$, and a single random configuration of the scatterer positions.
The dotted line is the Airy approximation~\eqref{eq:airy-total-cross-section}.}%
\label{fig:total-cross-section-airy}%
\end{figure}%
\par The total cross section numerically computed with Eq.~\eqref{eq:multi-total-cross-section-2} is shown in Fig.~\ref{fig:total-cross-section-airy}, and compared to the estimate of Eq.~\eqref{eq:airy-total-cross-section}.
The total cross section displays a long plateau whose height is approximately given by Eq.~\eqref{eq:airy-total-cross-section}.
The peak at $k=0$ is due to the singular behavior of the single-scatterer cross section in 2D, as shown in Fig.~\ref{fig:point-cross-section}, and the decrease for $k\gtrsim 17\,\varsigma^{-1}$ to the cancellation at $k=j_0/\alpha\simeq 24\,\varsigma^{-1}$.
The fluctuations comes from the fact that only one configuration of the scatterers is considered.
They can be explained by the existence of underlying resonances due to the multiple scattering of the particle.
We will study in detail the distribution of these resonances from the perspective of the complex plane of $k$ in our companion paper~\cite{GaspardD2022b}.

\section{Conclusions}\label{sec:conclusions}
We began this paper with an extension of elastic scattering theory to an arbitrary number of spatial dimensions, including dimension one.
Among the general concepts that we presented, there are the Green function, the cross section, the optical theorem, and the Airy diffraction pattern.
On top of that, we derived a scattering model for point scatterers which is expressed in terms of the scattering length $\alpha$, which is a universal parameter for low-energy scattering~\cite{Newton1982, Joachain1979, Taylor2006, Akkermans2007, Bolle1984b, Verhaar1985, Jeszenszki2018, GaspardD2018a, RamirezSuarez2013, Baye2000a}.
The latter quantity is explicitly related to the interaction potential through the wave function calculated in the inner region.
\par Then, we established the Foldy-Lax multiple scattering equations of a quantum particle in a random Lorentz gas of fixed point scatterers~\cite{Foldy1945, Lax1951, *Lax1952, Waterman1961, HuangK2010, Martin2018, Mishchenko2006}.
The problem is completely described by the symmetric but non-Hermitian multiple scattering matrix $\matr{M}(k)$.
We verified that the equations satisfy the optical theorem, and thus probability conservation, exploiting the properties of the free-space Green function.
In this regard, the optical theorem has the important consequence that the eigenvalues of $\matr{M}(k)$ have positive imaginary parts for $k\in\mathbb{R}$.
We also derived a position-space scattering matrix $\matr{S}(k)$.
We proved that this matrix is not unitary in the usual way, due to the non-normality of $\matr{M}(k)$, but nevertheless has unit-norm eigenvalues.
\par Furthermore, we computed the differential cross section of the random Lorentz gas for an incident plane wave.
We considered the ballistic and diffusive regimes.
In the ballistic regime, we showed that the differential cross section averaged over the scatterer configurations can be described by the Born approximation.
We also measured the angular fluctuations of the differential cross section for individual random configurations.
In the diffusive regime, a distinct Airy pattern is visible in the forward direction, meaning that the gas is mostly opaque for the incident wave.
In addition, this observation shows that the total scattering cross section of the gas is twice its geometrical cross section.
This is a manifestation of the extinction paradox, a famous paradox in wave mechanics~\cite{Massey1933, Brillouin1949, Hulst1981, Nussenzveig1992, Newton1982, Born2019, Bienaime2014}.
\par Finally, in this paper, we gave a general introduction on the random Lorentz gas model in arbitrary dimension.
This work will serve as a starting point for a more advanced study of the random Lorentz gas in the complex plane of the wavenumber which takes place in our companion paper~\cite{GaspardD2022b}.

\begin{acknowledgments}%
D.G. is grateful to Prof.~Pierre Gaspard for the review of the manuscript and the useful suggestions.
D.G. holds a Research Fellow (ASP - Aspirant) fellowship from the Belgian National Fund for Scientific Research (F.R.S.-FNRS).
This work was also supported by the F.R.S.-FNRS as part of the Institut Interuniversitaire des Sciences Nucléaires (IISN) under Grant Number 4.45.10.08.
\end{acknowledgments}%

\end{document}